\documentstyle[aps,twocolumn]{revtex}
\input epsf
\tightenlines
\begin{document}
\draft
\preprint{VECC/NTH97xxx}
\title{\bf Propagation of Charm Quarks in Equilibrating Quark-Gluon 
Plasma}
\author{Munshi Golam Mustafa, Dipali Pal, Dinesh Kumar Srivastava}
\address{Variable Energy Cyclotron Centre, 1/AF Bidhan Nagar, Calcutta 
700 064, India}
\date{\today}
\maketitle

\begin{abstract}
We study the propagation of charm quarks, produced 
from the initial fusion of partons,  in an equilibrating 
quark-gluon plasma which may be formed in the
wake of relativistic collisions of gold nuclei.
Initial conditions are taken from a self screened parton cascade model
and the chemical equilibration is assumed to proceed via
gluon multiplication and quark production.
The energy loss suffered by the charm quarks is obtained by evaluating the
drag force generated by the scattering with quarks and gluons in the medium.
We find that the charm quarks may loose only about 10\% of their initial
energy in conditions likely to be attained at the Relativistic
Heavy Ion Collider, while they may loose up to 40\% of their energy while
propagating in the plasma created at Large Hadron Collider.
We discuss the implications for signals of quark gluon plasma.

\end{abstract}
\narrowtext

\section{\bf INTRODUCTION}

Relativistic heavy ion collisions are being studied with the intention of 
investigating the properties of ultradense, strongly interacting matter-
quark-gluon plasma (QGP) \cite{HM96}. The last several years have witnessed
extensive theoretical efforts towards modeling these collisions. These studies
suggest that at collider energies, one may visualize the nuclei
as two clouds of valence and 
sea partons which pass through each other and interact~\cite{klaus}. 
The resulting partonic interactions may then produce a dense plasma
of quarks and gluons. This plasma will expand and become cooler 
and more dilute. 
If quantum chromodynamics admits a first-order deconfinement or chiral 
phase transition, this plasma will pass through a mixed phase of quarks, 
gluons, and hadrons, before the hadrons loose thermal contact and stream 
freely towards the detectors.  

One would like to understand several aspects of this evolution, viz., 
how does the initial partonic system evolve and how quickly does it attain
kinetic equilibrium? How quickly, if at all, 
does it attain chemical equilibrium?
And finally, how can we uncover the history of this evolution
by studying the spectra of the produced particles, many of which may 
decouple from the interacting system only towards the end? These and related
questions have been actively debated in recent times.
Thus, it is believed by now that the large initial parton density  may force
many collisions among the partons in a very short time and lead to a kinetic
equilibrium~\cite{kin}. The question of chemical equilibration is more
involved, as it depends on the time available to the system. If the time
available is too short (3--5 fm/c), as at the
energies ($\sqrt{s} \leq 100$ GeV/nucleon) accessible at the
Relativistic Heavy Ion Collider (RHIC), the QGP will end its journey far
away from chemical equilibrium. At the energies ($\sqrt{s}
\leq 3$ TeV/nucleon) that will be achieved at the CERN Large Hadron
Collider (LHC) this time could be large (more than 10 fm/$c$), driving
the system very close to chemical equilibration~\cite{biro,chem,kg,ja,xmh},
if there is only a longitudinal expansion.
However, this time is also large enough to enable a rarefaction wave from
the surface of the plasma to propagate to the center 
($ \tau_s \sim R_T/c_s$; $R_T$ is the transverse nuclear dimension and
$c_s$ is the speed of sound),
 thus introducing large transverse velocity (gradients) in the system. 
The large transverse velocities may
impede the chemical equilibration by introducing a faster cooling. 
The large velocity gradients may drive the system away from chemical
equilibration by introducing an additional source of depletion of the
number of  partons~\cite{munshi} in a given fluid element.

Can we identify a probe for these rich details of the evolution? Dileptons
and single photons have long been considered as useful probes of the various
stages of the plasma as once produced they hardly ever interact and
thus carry the details of the circumstances of their production.  They are
however produced at every stage of the collision and in an expanding
system their number is obtained by an integration over the four-volume
of the interaction zone. This moderates our estimate of the efficacy of these
signals in several competing ways, as the space-time occupied by the hot
plasma is small.

 Consider the case of single photons.
At very early times the temperature is rather high and we have a copious
production of photons having a large transverse momentum. This should give us
a reliable information about the initial conditions of the plasma.
 However, the transverse flow of the system is very moderate at early times.
By the time the flow and other aspects of the QGP develop,
 the temperature would have dropped considerably and we have a large
production of photons having  smaller transverse momenta. However, the
photons having a small transverse momentum will also be produced,
and in much larger numbers, from the
hadronic reactions and decays of vector mesons in the hadronic matter. Thus
disentangling the development of the later stages of the
evolution in the QGP would entail
a very detailed understanding of the contributions of the hadronic
matter. This is not trivial. 

Next consider dileptons. Again, large mass
dileptons, having their origin in early times, will be only marginally
affected by these developments, e.g, the flow. 
The low-mass dileptons produced at later times will  be affected
strongly by the flow. However they will have, in addition to the
contributions from the plasma, a large contribution from hadronic
reactions. We shall be required to understand them in detail before
we can confidently embark upon the task of deciphering the, more involved,
development of the QGP,  as it gets cool (see, e.g.,Ref,~\cite{munshi}). At the
same time, there have been suggestions that correlated charm or bottom decay
may give a large background to the dileptons from the QGP. 
We shall come back to this aspect later.

Why should we be so concerned about these late developments in the plasma?
Firstly, it is not difficult to imagine that the detailed constitution of
the hadronic matter should somehow reflect the state of chemical equilibration
of the QGP at the time of the phase transition. Secondly, the transverse
velocities, which may develop during the QGP phase will remain 
unaltered during the mixed phase, if such a phase is attained.
The development of the transverse velocities should also  be affected by
the speed of sound, which may change rapidly
as we approach the transition temperature. In  the absence of transverse
 velocities, the mixed phase will last for a very long period of 
time~\cite{vesa}. And
lastly, it is not at all clear that a non-equilibrated plasma will 
be adiabatically converted to hadronic matter, a scheme which is
often invoked in hydrodynamic descriptions of the system.

In the present work we report on an investigation of the energy loss
suffered by charm quarks in a chemically equilibrating
 plasma. This study has several interesting features.
Firstly, heavy quarks are mostly produced from the initial fusion of partons
in the colliding nucleons. If the initial temperature is high, at least
charm quarks may also be produced at very early times
(mostly from $gg\,\rightarrow\,c\overline{c}$, but also from
$q\overline{q}\,\rightarrow\,c\overline{c}$). There is no production
of charm quarks at later times and none in the hadronic matter. This makes
them a good candidate for a probe of QGP.
Again, as their number is not large, one can confidently ignore the reverse
process ($c\overline{c}\,\rightarrow\,gg\,{\mathrm {or}} \, q\overline{q}$),
where $q$ denotes one of the lighter quarks. 
(Of-course the associated production/suppression of $J/\Psi$ and $\Upsilon$
is a subject by itself.) The number of initially produced
charm quarks  (from fusion of primary partons) can be evaluated with a
great degree of confidence using the perturbative QCD~\cite{ramona}. 
The thermally produced heavy
quarks can also be estimated in terms of the temperatures and the fugacities
during the early stage of evolution, when they are produced. Thus, the total
number of c or b quarks gets frozen very early in the history of the
collision. This can be of great help, as we are then left with the
task of determining their $p_T$ distribution, whose details
may reflect the developments in the plasma, but  which is normalized
to a given number of c (or b) quarks. 

Now consider the central slice ($y=0$) of a collision which leads to 
production of a charm quark
pair from initial fusion and QGP. The charm quarks will be produced on the
order of a time scale $1/2m_c\,\simeq\,0.07$ fm/c, while the bottom quarks
will similarly be produced over a time of $\sim 0.02$ fm/c. Thus, immediately 
upon production, these quarks will find themselves in a deconfined matter
 which is rapidly evolving, first towards kinetic equilibration and then
towards chemical equilibration. The heavy quarks, as well as the light quarks 
and gluons will undergo scattering, with an interesting difference. The
heavy quarks are not very  likely to have  a frequent  and drastic change in
their direction upon bombardment  by massless quarks and gluons.
These collisions will amount to a drag force.
Svetitsky~\cite{ben} has argued that this drag force is
rather large and it will bring down the velocity of charm quarks
down to thermal velocity very quickly.

In the present work we show that when we account for the non-equilibrium
nature of the plasma, this drag force reduces considerably. In particular,
we find that the charm quarks produced at the RHIC may loose only up to 10\%
of their initial momentum though at the LHC they may loose up to 40\%
of their initial momentum during their passage through the QGP phase of the
matter.

The momenta of these quarks are likely to be reflected in the corresponding 
quantities of D mesons as the c quarks should
pick up a light quark, which are in great abundance and hadronize.
(One may add an interesting scenario, where the D mesons travel through
a string of QGP droplets  and loose further energy ~\cite{ben2}.)
 Finally, they would scatter with hadronic matter before decoupling.
Considering that $\sigma_{D\pi} \,\ll \, \sigma_{\pi \pi}$, e.g.,
 it is most likely that the heavy mesons would decouple quickly from the
hadronic phase. 

There has been an attempt~\cite{shur,lin} to estimate 
the change in the distribution of
 D mesons (open charm) by introducing a rate of energy loss
 $dE/dx \approx $ $-2$ GeV/fm for the charm and bottom quarks. We recall that
 this value was originally obtained~\cite{rud1} for very energetic
 quarks. For this reason, this estimate also derives a considerable contribution
 from the radiation of gluons. Charm quarks are very {\em massive}, and thus
 the radiation mechanism is absent~\cite{rud2} unless their energy is
  exceptionally high.
  We shall
 be concerned with charm quarks having a $p_T$ of a few GeV/c, in our study.
For these energies, the drag force will come mostly from scatterings with
light quarks and gluons~\cite{ben} in a formulation which is equivalent 
to the treatment of Braaten and Thoma~\cite{markus} for this purpose.
 We show in the present work that this energy loss can
be very low for a chemically non-equilibrated plasma.

In the next Section, we  briefly recall the initial conditions,
and  hydrodynamic and chemical
evolution of the plasma in a (1+1) dimensional expansion.
In Section III we give the  formulation for drag force operating on the 
charm quarks in such an equilibrating plasma. We discuss our results in 
section IV. Finally, in section V we give a brief summary.

\section{\bf INITIAL CONDITIONS, HYDRODYNAMIC EXPANSION,
AND CHEMICAL EQUILIBRATION}

It is quite clear that the fate of  the heavy quarks will
 depend on the initial
conditions and the history of evolution of the plasma.
Fortunately, by now,
a considerable progress has been achieved in our understanding of 
the parton cascades which develop in the wake of the collisions. 
 Thus, in the recently formulated self-screened parton
cascade~\cite{sspc} model early hard scatterings produce a medium which
 screens the longer
ranged color fields associated with softer interactions. When two
heavy nuclei collide at sufficiently high energy, the screening occurs
on a length scale where perturbative QCD still applies.
This approach yields predictions for the initial conditions of the
forming QGP without the need for any {\it ad-hoc} momentum and virtuality
cut-off parameters~\cite{klaus}. These calculations also show that the QGP
likely to be formed in such collisions could be very hot and initially
far from chemical equilibrium. 

We assume that kinetic equilibrium has been achieved when the momenta of
partons become locally isotropic. At the collider energies it has been 
estimated that, $\tau_i\approx 0.2-0.3$ fm/$c$~\cite{kin}. 
 Beyond this point, further expansion is described by
 hydrodynamic equations and the chemical equilibration is governed by 
a set of master equations which are driven by the two-body reactions
($gg\,\leftrightarrow\,q\bar{q}$) and gluon multiplication and its
inverse process, gluon fusion ($gg\,\leftrightarrow\,ggg$). The other
(elastic) scatterings help maintain thermal equilibrium.
The hot matter continues to expand and cools due to expansion and
chemical equilibration.  We shall somewhat arbitrarily terminate
the evolution once the energy density reaches some critical
value (here taken as $\epsilon_f=1.45$ GeV/fm$^3$ \cite{com}). 

The expansion of the system is described by the equation for 
conservation of energy and momentum of an ideal fluid:
\begin{equation}
\partial_\mu T^{\mu \nu}=0 \; , \qquad
 T^{\mu \nu}=(\epsilon+P) u^\mu u^\nu + P g^{\mu \nu} \, ,
\label{hydro}
\end{equation}
where $\epsilon$ is the energy density and $P$ is the pressure measured 
in the frame comoving with the fluid. 
The four-velocity vector $u^\mu$ of the 
fluid satisfies the constraint $u^2=-1$. 
For a partially equilibrating plasma the distribution functions for 
gluons and  quarks can be scaled through equilibrium distributions as
\begin{equation}
g_{i}(q,T,\lambda_i)= \lambda_{i} g^{\mathrm{eq}}_i(q,T) \ \ ,
 \label{befd}
\end{equation}
where $g^{\mathrm{eq}}_i(q,T)=({e^{\beta{u\cdot q}}\mp 1})^{-1}$
is the BE (FD) distributions for gluons
(quarks), and $\lambda_i$ is the fugacity for parton species $i$, which
describes the deviation from chemical equilibrium. This fugacity factor
takes into account undersaturation of parton phase space density,
$i.e.$, $0\leq\lambda_i\leq 1$.
The equation of state for a partially equilibrated plasma of massless 
particles can be written as
\cite{biro}
\begin{equation}
\epsilon=3P=\left[a_2 \lambda_g +  b_2 \left (\lambda_q+\lambda_{\bar q}
\right ) \right] T^4 \, ,
\label{eos}
\end{equation}
where $a_2=8\pi^2/15$, $b_2=7\pi^2 N_f/40$, $N_f \approx 2.5$ is 
the number of dynamical quark flavors. Now, the density of an
equilibrating partonic system can be written as 
\begin{equation}
n_g=\lambda_g \tilde{n}_g,\qquad 
 n_q=\lambda_q \tilde{n}_q,
\end{equation}
where $\tilde{n}_k$ is the equilibrium density for the parton species $k$:
\begin{equation}
\tilde{n}_g=\frac{16}{\pi^2}\zeta(3) T^3=a_1 T^3,
\end{equation}
\begin{equation}
\tilde{n}_q=\frac{9}{2\pi^2}\zeta(3) N_f T^3=b_1 T^3.
\end{equation}
We further assume that $\lambda_q=\lambda_{\bar{q}}$.  The equation of
state (\ref{eos}) implies the speed of sound $c_s=1/\sqrt{3}$.

The master equations \cite{biro} for the dominant chemical reactions 
$gg \leftrightarrow ggg$ and $gg \leftrightarrow q\bar{q}$  are
\begin{eqnarray}
\partial_\mu (n_g u^\mu)&=&n_g(R_{2 \rightarrow 3} -R_{3 \rightarrow 2})
                    - (n_g R_{g \rightarrow q}
                       -n_q R_{q \rightarrow g} ) \, , \nonumber\\
\partial_\mu (n_q u^\mu)&=&\partial_\mu (n_{\bar{q}} u^\mu)
                     = n_g R_{g \rightarrow q}
                       -n_q R_{q \rightarrow g},
\label{master1}
\end{eqnarray}
in an obvious notation. 

If we assume the system to undergo  a purely longitudinal boost invariant 
expansion, (\ref{hydro}) reduces to the well known relation \cite{bj}
\begin{equation}
\frac{d\epsilon}{d\tau}+\frac{\epsilon+P}{\tau}=0,
\label{long}
\end{equation}
where $\tau$ is the proper time. This equation implies
\begin{equation}
\epsilon\, \tau^{4/3}=\,{\mathrm {const}.}
\label{epstau}
\end{equation}
and the chemical master equations reduce to \cite{biro}
\begin{eqnarray}
\frac{1}{\lambda_g}\frac{d \lambda_g}{d\tau}
+\frac{3}{T}\frac{dT}{d\tau} +
\frac{1}{\tau} 
 &=&
R_3 ( 1- \lambda_g ) -2 R_2 \left( 1-\frac{\lambda_q \lambda_{\bar{q}}}
{\lambda_g^2}\right) \, , 
\nonumber\\
\frac{1}{\lambda_q}\frac{d \lambda_q}{d\tau}
+\frac{3}{T}\frac{dT}{d\tau} +
\frac{1}{\tau} 
 &=&
R_2 \frac{a_1}{b_1} \left(
\frac{\lambda_g}{\lambda_q}-\frac{\lambda_{\bar{q}}}{\lambda_g}\right)\, ,
\label{master_long}
\end{eqnarray}
which are then solved numerically for the fugacities with 
the initial conditions obtained from self screened parton cascade model
\cite{sspc}.
We quote the rate constants $R_2$ and $R_3$ appearing in Eq.~(\ref{master1})
 for the sake of completeness~\cite{biro};
\begin{eqnarray}
R_2 & \approx & 0.24 N_f \alpha_s^2 \lambda_g T \ln (1.65/\alpha_s \lambda_g),
\nonumber\\
R_3 & = & 1.2 \alpha_s^2 T (2\lambda_g-\lambda_g^2)^{1/2},
\end{eqnarray}
where the color Debye screening and the Landau - Pomeranchuk - Migdal effect
suppressing the induced gluon radiation have been taken into account,
explicitly.

\section{DIFFUSION OF HEAVY QUARKS IN QUARK GLUON PLASMA}

The early results of parton cascade model~\cite{klaus} do indicate that
the charm quarks which are produced from initial fusion in heavy ion
collisions may have a transverse momentum distribution given by a power
law, initially. With the passage of time, these distributions 
start evolving 
into an exponential shape, due to scattering with other partons, which
is characteristic of  a thermalized system of particles. Will these
heavy quarks stay in thermal equilibrium, till the end of the QGP
phase? For this we have to continue the cascading till the temperatures
have dropped to $T_c$ or the energy density has become too low.
The parton cascade model also includes all the effects like scatterings
and radiation, if any, on the heavy quark. We shall
report on such an effort in a future publication~\cite{klaus_charm}.

 In the present work, we adopt the procedure developed by Svetitsky
\cite{ben},
where one visualizes the effect of partonic collisions as leading to a drag
force. We first generalize the
treatment of Svetitsky to a non-equilibrated plasma and then evaluate
the time variation of the so-called drag and the diffusion coefficients,
to see whether heavy quarks will actually stop and diffuse at RHIC and
LHC energies.

We write the Boltzmann equation for the density $f({\mathbf p}, t)$ for
a heavy quark in phase space:
\begin{equation}
\frac{\partial}{\partial t}f({\mathbf p},t)=\left[\frac{\partial f}
{\partial t}\right]_{\mathrm {collisions}} \,\, . \label{boltz}
\end{equation}

It is assumed that there is no external force acting on the heavy
quark and that the phase space distribution $f$ does not depend on the
position of the quark. The right hand side of Eq.~(\ref{boltz})
represents a collision integral  given by,
\begin{eqnarray}
R({\mathbf p},t)&=& \left [\frac{\partial f}{\partial t} \right
]_{\mathrm{collisions}} \ \ \nonumber \\
&=& \int d^3k \left [ w({\mathbf{p+k,k}})f({\mathbf{p+k}})
-w({\mathbf{p,k}})f({\mathbf p})\right ], \nonumber \\
\label{colrate}
\end{eqnarray}
where $w({\mathbf{p,k}})$ is the rate of collisions which 
change the momentum of the charmed quark from $\mathbf p$ to 
${\mathbf {p-k}}$.  The Eq.({\ref{colrate}) can be written
explicitly~\cite{ben} as 
\begin{eqnarray}
R({\mathbf p},t)&=& {\frac{1}{2E_{\mathbf p}}}\int {\frac{d^3{\mathbf q}}
{(2\pi)^32E_{\mathbf q}}} 
\int{\frac{d^3{\mathbf q}'}{(2\pi)^3 2E_{{\mathbf q}'}}} 
\int{\frac{d^3{\mathbf p}'} {(2\pi)^32E_{{\mathbf p}'}}} \nonumber \\
&\times& {\frac{1}{\gamma_c}} \sum |{\cal M}|^2
(2\pi)^4 \delta^4\left (p+q-p'-q' \right ) \nonumber \\
&\times& \left [ f({\mathbf p'})g({\mathbf q'}){\tilde {g}}({\mathbf q})-
f({\mathbf p})g({\mathbf q}){\tilde {g}}({\mathbf q'}) \right ] ,
\label{exprate} 
\end{eqnarray}
where ${\mathbf{p'=p-k}}$, ${\mathbf{q'=q+k}}$, and $\gamma_{c}$ is spin
and color degeneracy of the charm quarks. In the above, $g$ represents
the distribution for quarks or gluons approximated as Eq.~(\ref{befd}), earlier.
 We have also introduced 
the Bose enhancement (Pauli  suppression) factors,
 $\tilde {g}(q)=1\pm g(q)$ for gluons
( quarks) ~\cite{daphne_thesis}. The matrix elements $|{\cal M}|^2$ 
for elastic processes can be obtained from~\cite{combridge,ben}.

We may further simplify the above by
assuming soft scatterings according to Landau~\cite{landau} to obtain,
\begin{equation}
R\approx \int d^3{\mathbf{k}} \left [ {\mathbf k}\cdot
{\frac{\partial}{\partial {\mathbf p}}}+
{\frac{1}{2}}k_ik_j{\frac{\partial^2}{{\partial {p}_i}{\partial
{p}_j}}} \right ] (wf) . \label{landrate}
\end{equation}

Now Eq.~(\ref{boltz}) reduces to well known Fokker-Planck form
\begin{equation}
{\frac{\partial f}{\partial t}}={\frac{\partial}{\partial {p_i}}}\left
[A_i({\mathbf p})f +
{\frac{\partial}{\partial{p_j}}}\left(B_{ij}({\mathbf p})f
\right)\right] , \label {fokker}
\end{equation}
where the kernels
\begin{eqnarray}
A_i& =& \int d^3{\mathbf k} w({\mathbf{p,k}})k_i  \ \ , \label{drag}\\
B_{ij}& =& {\frac{1}{2}} \int d^3{\mathbf k} w({\mathbf {p,k}})k_ik_j \ \ ,
\label{diffuse}
\end{eqnarray}
can be identified with the drag and diffusion 
coefficients, respectively, using the 
Langevin formalism \cite{ben}.
 In particular, $A_{i}$ and $B_{ij}$
are given by,
\begin{eqnarray}
A_{i}&=& {\frac{1}{2E_{\mathbf p}}}\int {\frac{d^3{\mathbf q}}
{(2\pi)^32E_{\mathbf q}}} 
\int{\frac{d^3{\mathbf q}'}{(2\pi)^3 2E_{{\mathbf q}'}}} 
\int{\frac{d^3{\mathbf p}'} {(2\pi)^32E_{{\mathbf p}'}}} \nonumber \\
&\times& {\frac{1}{\gamma_c}} \sum |{\cal M}|^2
(2\pi)^4 \delta^4\left (p+q-p'-q' \right ) \nonumber \\
&\times& g({\mathbf q}){\tilde {g}}({\mathbf q'}) \left [
\left(p-p'\right )_i \right ] \nonumber \\
&\equiv & \langle \! \langle \left (p-p'\right)_i 
         \rangle \! \rangle  \ \ , \label{drag1}\\
B_{ij}&=&{\frac{1}{2}}\langle \! \langle\left (p'-p\right)_i\left(p'-p\right)_j
 \rangle \! \rangle \ \ .
\label{diffuse1}
\end{eqnarray}
We note that the Eqs.~(\ref{drag1}) and (\ref{diffuse1}) depend only on 
${\mathbf p}$, and thus we may write~\cite{ben};
\begin{eqnarray}
A_i&=& p_i A(p^2) \ \ , \label{drag2}\\
B_{ij}& =& \left[ \delta_{ij} - {\frac{p_ip_j}{p^2}}\right
]B_0(p^2)+{\frac{p_ip_j}{p^2}}B_1(p^2)\ \ ,\label{diffuse2}\\
\end{eqnarray}
where
\begin{eqnarray}
A&=& 
\langle \! \langle 1 \rangle \! \rangle
 - \langle \! \langle{\mathbf {p\cdot p'}}\rangle \! \rangle/p^2 \ \ ,
  \label{drag3}\\
B_0&=& {\frac{1}{4}} \left[ \langle \! \langle{p'}^2
\rangle \! \rangle - \langle \! \langle ({\mathbf{p'\cdot p}})^2
\rangle \! \rangle/p^2
\right ] \ \ , \label{diffuse3} \\
B_1&=& {\frac{1}{2}}  \left [ \langle \! \langle ({\mathbf{p' \cdot p}})^2
\rangle \! \rangle/p^2 - 2 \langle \! \langle
{\mathbf{p' \cdot p}} \rangle \! \rangle 
+ p^2 \langle \! \langle 1\rangle \! \rangle \right ].
\label{diffuse4}
\end{eqnarray}
The integrals appearing in the equations given above can be further
simplified, by solving the kinematics in the center of mass
frame of the colliding particles, so that we can write,
\begin{eqnarray}
\langle \! \langle F(\mathbf{p'})\rangle \! \rangle &=& {\frac{1}{512\pi^4\gamma_c}}
{\frac{1}{E_{\mathbf p}}} \int_0^\infty q \ dq \int_{-1}^1 d(\cos \chi)
\nonumber \\
&\times&{\frac{\sqrt{(s+m_c^2-m_{g(q)}^2)^2-4sm_c^2}}{s}} \ 
g(E_{\mathbf q}) \nonumber \\
&\times& \int_{-1}^1 d\cos{\theta_{\mathrm{c.m.}}} \sum |{\cal M}|^2 
\int_0^{2\pi} d\phi_{\mathrm{c.m.}}\nonumber \\
&\times& e^{\beta E_{\mathbf q'}} g(E_{\mathbf q'}) F(\mathbf{p'}) \ .
\label{final} 
\end{eqnarray}
where 
and $s = (E_{\mathbf p}+E_{\mathbf q})^2-(\mathbf{p} +\mathbf{q})^2$,  
$E_{\mathbf{q'}} =E_{\mathbf {q}}+E_{\mathbf {p}}-E_{\mathbf{p'}}$ and
$\mathbf p'$ is a function of $\mathbf p$, $\mathbf q$ and
$\theta_{\mathrm{c.m.}}$.
 Note that in addition to other 
differences due to the introduction of the
 mass of quarks and gluons  (see later) and the quantum statistics,
 this expression
differs~\cite{thanks_ben} by a factor of 2 from the corresponding Eq.~(3.6)
 of Ref.\cite{ben}.
The quantum statistical correction for gluons employed here introduces a
divergence as $q\rightarrow 0$ which is absent if we
use a Boltzmann distribution for them.
We have avoided this divergence~\cite{daphne_thesis} by using the
thermal gluon mass;
\begin{equation}
m_g^2= \lambda_g\left (1+{\frac{N_f}{6}}\right ) {\frac {g^2T^2}{3}} \
\ , \label{gmass}
\end{equation}
where $g$ is QCD coupling constant ($\alpha_s=g^2/4\pi$). In order to 
retain a certain degree of self-consistency, we have also
used the thermal quark mass;
\begin{equation}
m_q^2= \left (\lambda_g+{\frac{\lambda_q}{2}} \right ) {\frac{g^2T^2}
{9}} \ \ . \label{qmass}
\end{equation}
We  thus correct these masses for the absence of chemical equilibrium.

We have also evaluated the integrals appearing in Eqs.~(\ref{drag1})
etc. by a direct Monte Carlo procedure, 
and the results were found to agree with the simplified expressions 
given above.

\section{RESULTS}

\subsection{Chemically Equilibrated Plasma}
As a first step we show in Fig.~1(a) the variation of the drag coefficient
$A$ with the momentum of the charm quark $p$ for a {\em chemically
equilibrated} QGP at a temperature of 500 MeV. First of all, we note that
the consequences of introducing the quantum statistics are quite small.
We must comment on the difference between our results
and those of Svetitsky~\cite{ben} before proceeding. Firstly, we have
used $\alpha_s=$ 0.3, instead of 0.6 used by Svetitsky. Secondly, we have used
$N_f$, the number of flavors, as 2.5 instead of 2. Finally, we have
have used the Debye screening mass 
 $\mu =\mu_D=\sqrt{4 \pi \alpha_s \lambda_g}\ T$ in the internal gluon
propagator in the $t-$ channel exchange diagram,
 to remain consistent with the description used while evaluating the
chemical evolution of the QGP. This is in addition to the numerical factor
of 2 mentioned above. We have verified that reverting to the values
used by Svetitsky, we reproduce the results of Ref.~\cite{ben}, but for
the numerical factor of 2.

\begin{figure}
\epsfxsize=3.25in
\epsfbox{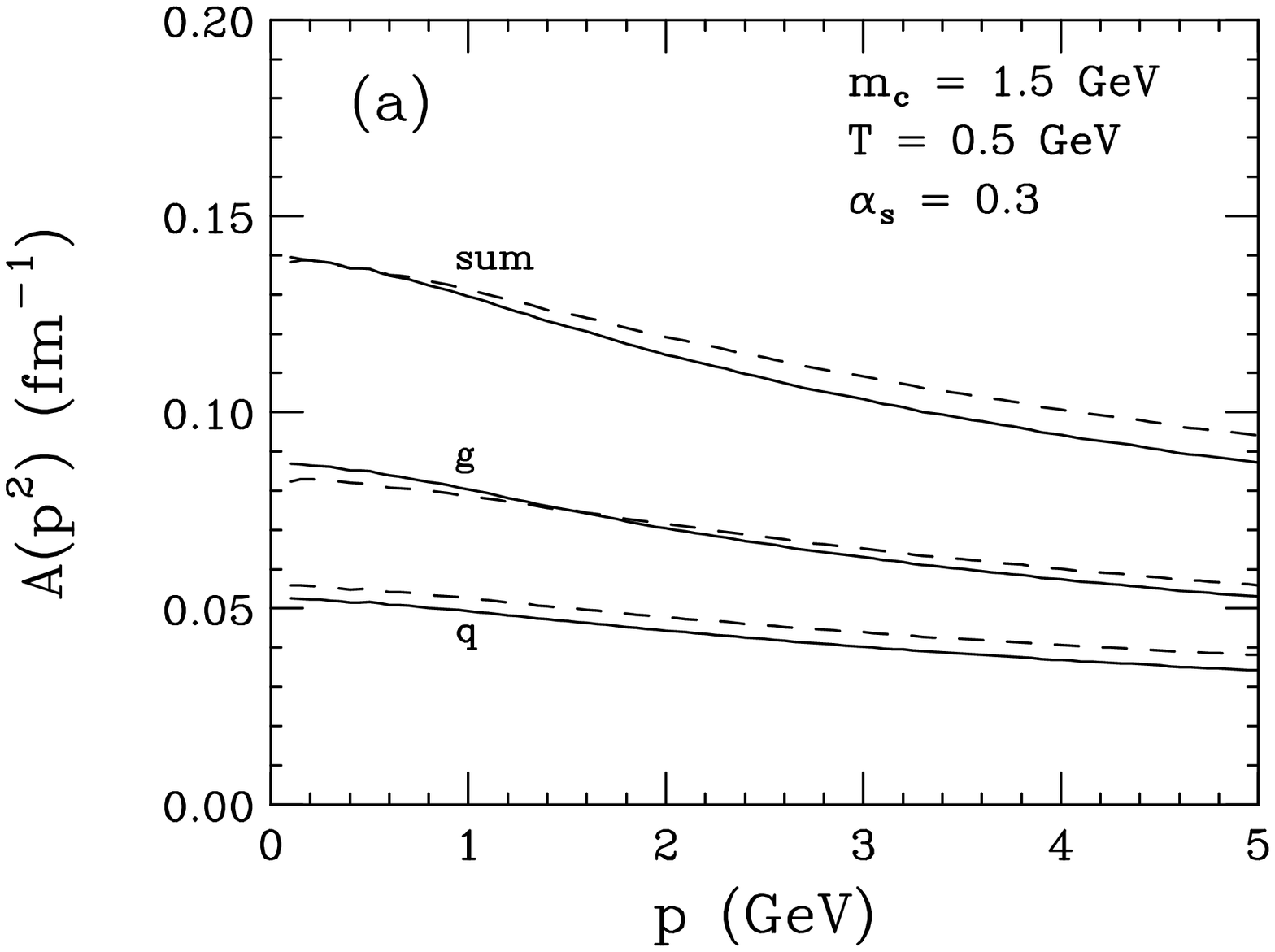}
\vskip 0.10in
\epsfxsize=3.25in
\epsfbox{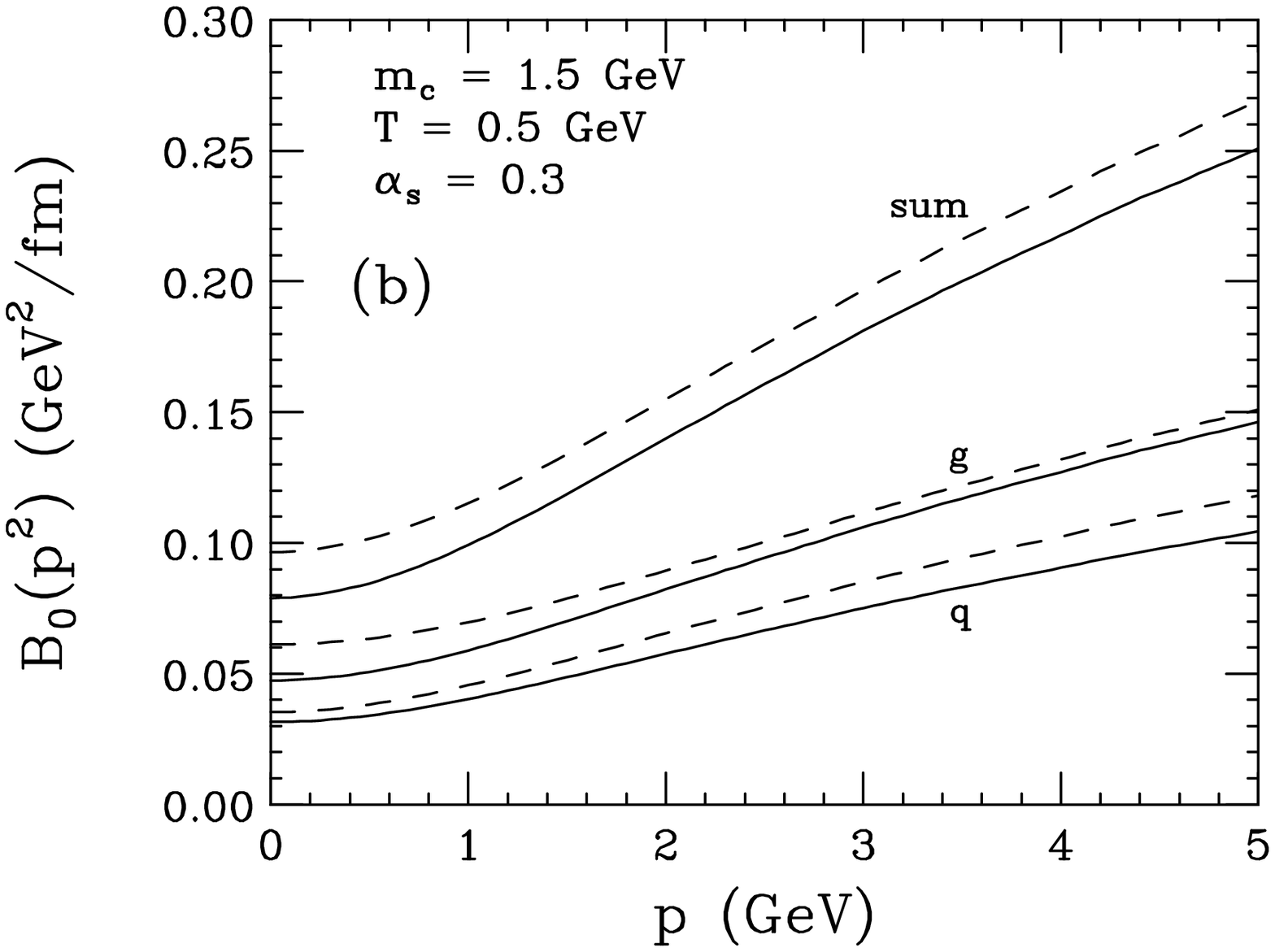}
\vskip 0.10in
\epsfxsize=3.25in
\epsfbox{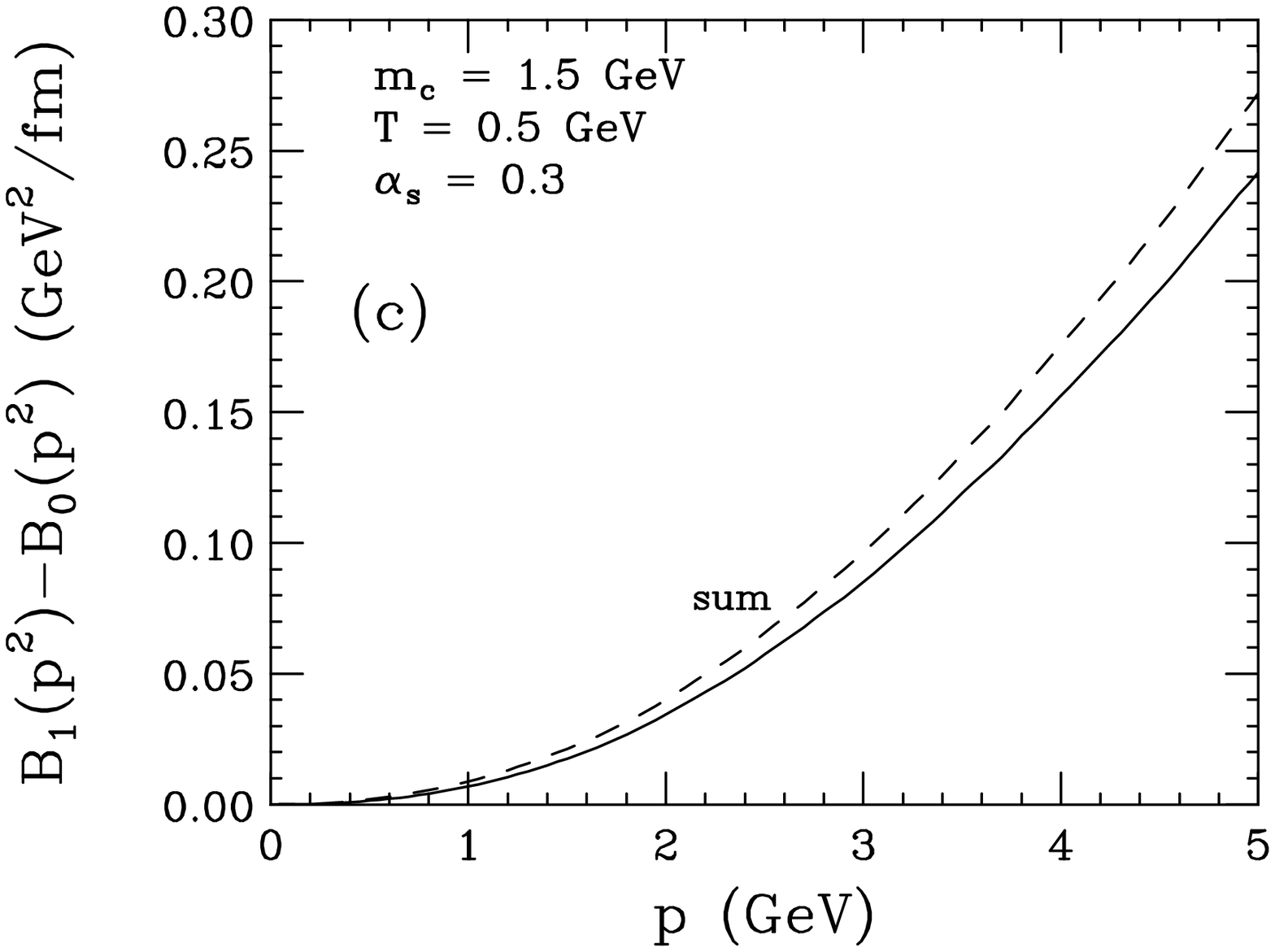}
\vskip 0.1in
\caption{The drag ($A$) and the diffusion ($B_0$ and $B_1-B_0$) coefficients 
for charm quarks in a fully equilibrated quark gluon plasma at a temperature 
of 500 MeV. The solid curves give the results with the proper quantum 
statistics while the dashed curves give the results with the Boltzmann 
distribution.}
\end{figure}

 We find that our result for the drag coefficient $A$ is a factor about
 three smaller than  obtained by Svetitsky. (Note that the results scale
 with $\alpha_s^2$.) This will have an important consequence as we shall 
 see later.

The momentum dependence of the diffusion coefficients $B_0$ and $B_1 -B_0$
are shown in Figs.~1(b)  \& 1(c). While the general behavior remains
similar to the results of Svetitsky, our values are smaller, due to
the reasons given earlier.
\begin{figure}
\epsfxsize=3.25in
\epsfbox{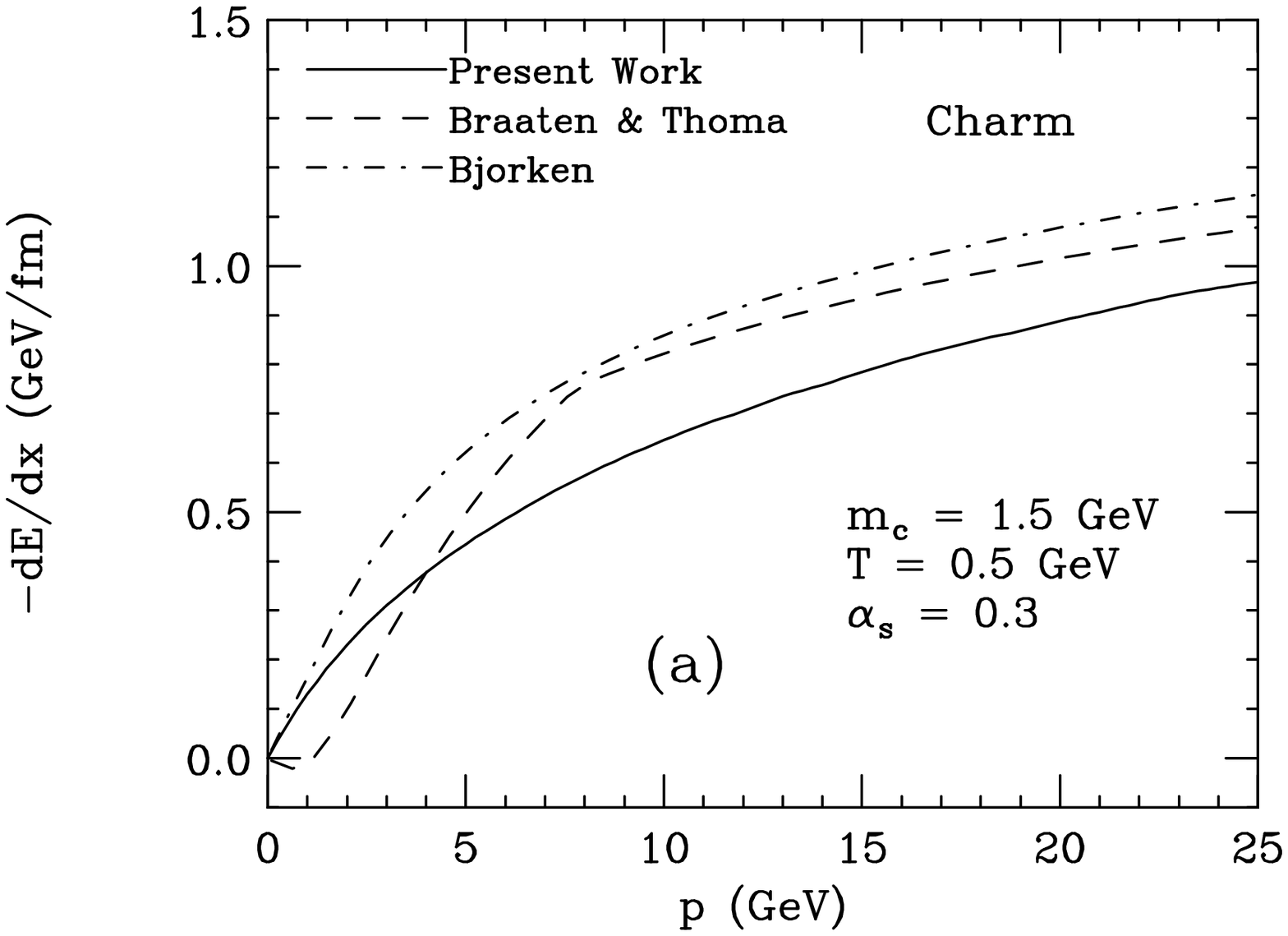}
\vskip 0.1in
\epsfxsize=3.25in
\epsfbox{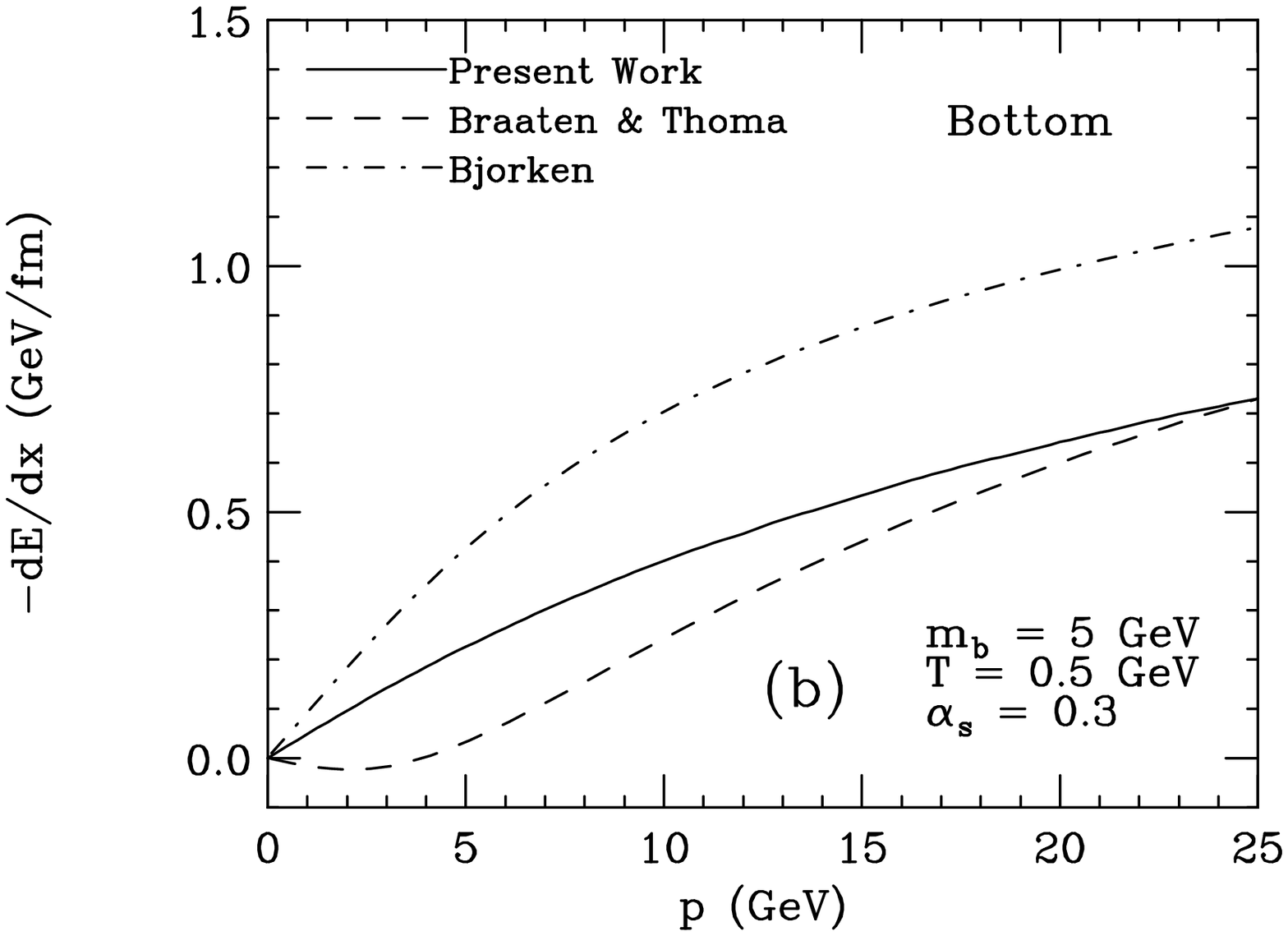}
\vskip 0.1in
\caption{The energy loss ($dE/dx$) of heavy quarks as a function of their 
momenta for $T$=0.5 GeV and $\alpha_s=0.3$.}
\end{figure}
The energy loss $dE/dx$ is related 
to the drag coefficient $A$;
\begin{equation}
\frac{dE}{dx}=-A(p^2) p \ . \label{dedx}
\end{equation}
In Fig.~2 we have compared our results for the energy loss suffered by
charm and bottom quarks with the results of Braaten and Thoma~\cite{markus}
and Bjorken~\cite{bjorken} (adopted to the case of heavy quarks). Our results
agree with those of Braaten and Thoma within 10\%. We note that even a fully
equilibrated plasma, whose temperature is kept fixed at 500 MeV, a charm
quark having a momentum of 1--5 GeV will travel for almost 8--10 fm before
coming to rest.
The results would be completely different for a $dE/dx$ of $\approx$
$-$2 GeV/fm assumed in the studies reported in Ref.~\cite{shur,lin}. We feel
however that at exceptionally high energies, the additional mechanism of gluon
radiation may enhance the energy loss. That should be of interest if we
wish to study the propagation of a heavy quark jet in the QGP. 

\begin{figure}
\epsfxsize=3.25in
\epsfbox{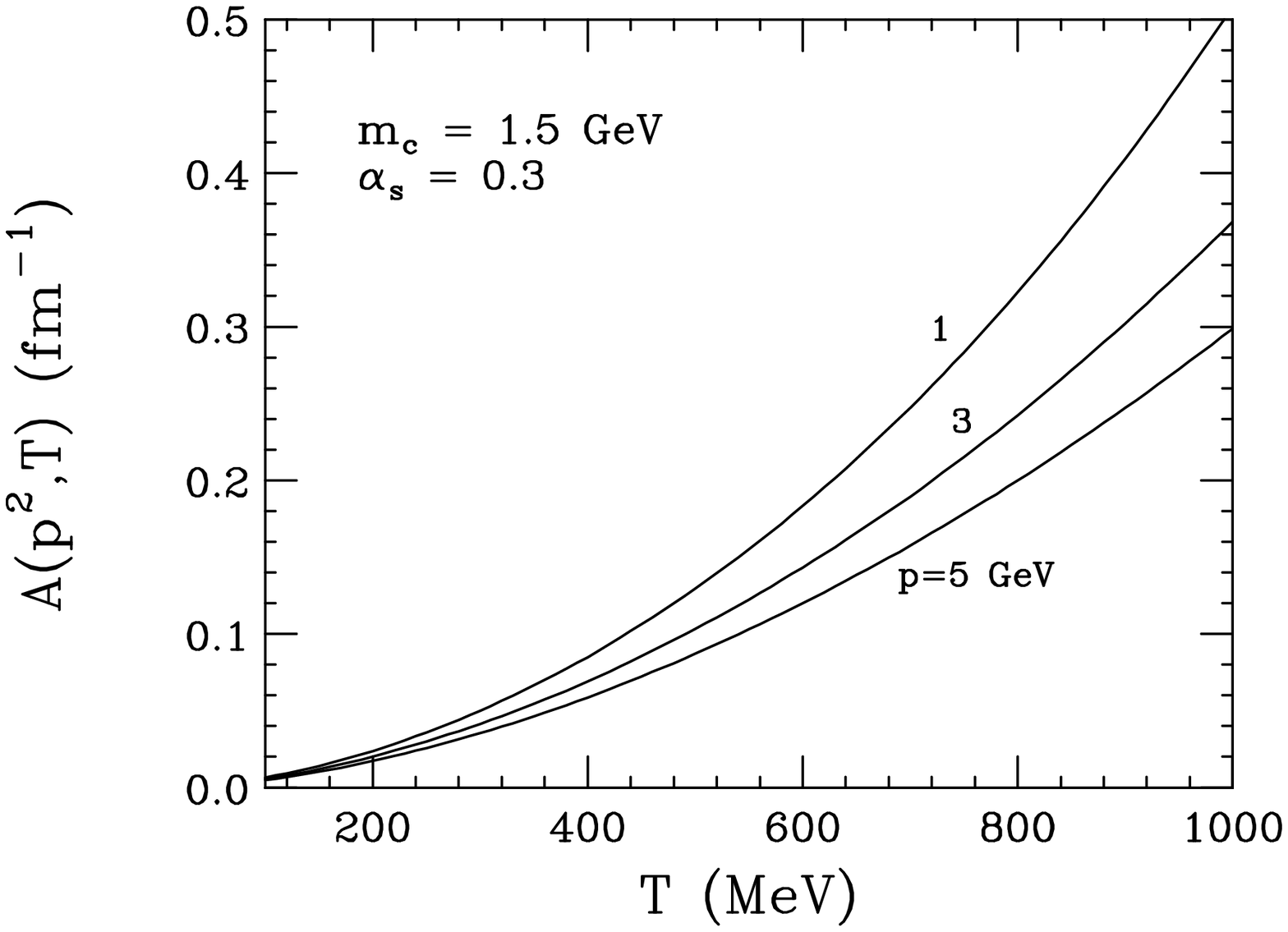}
\vskip 0.1in
\epsfxsize=3.25in
\epsfbox{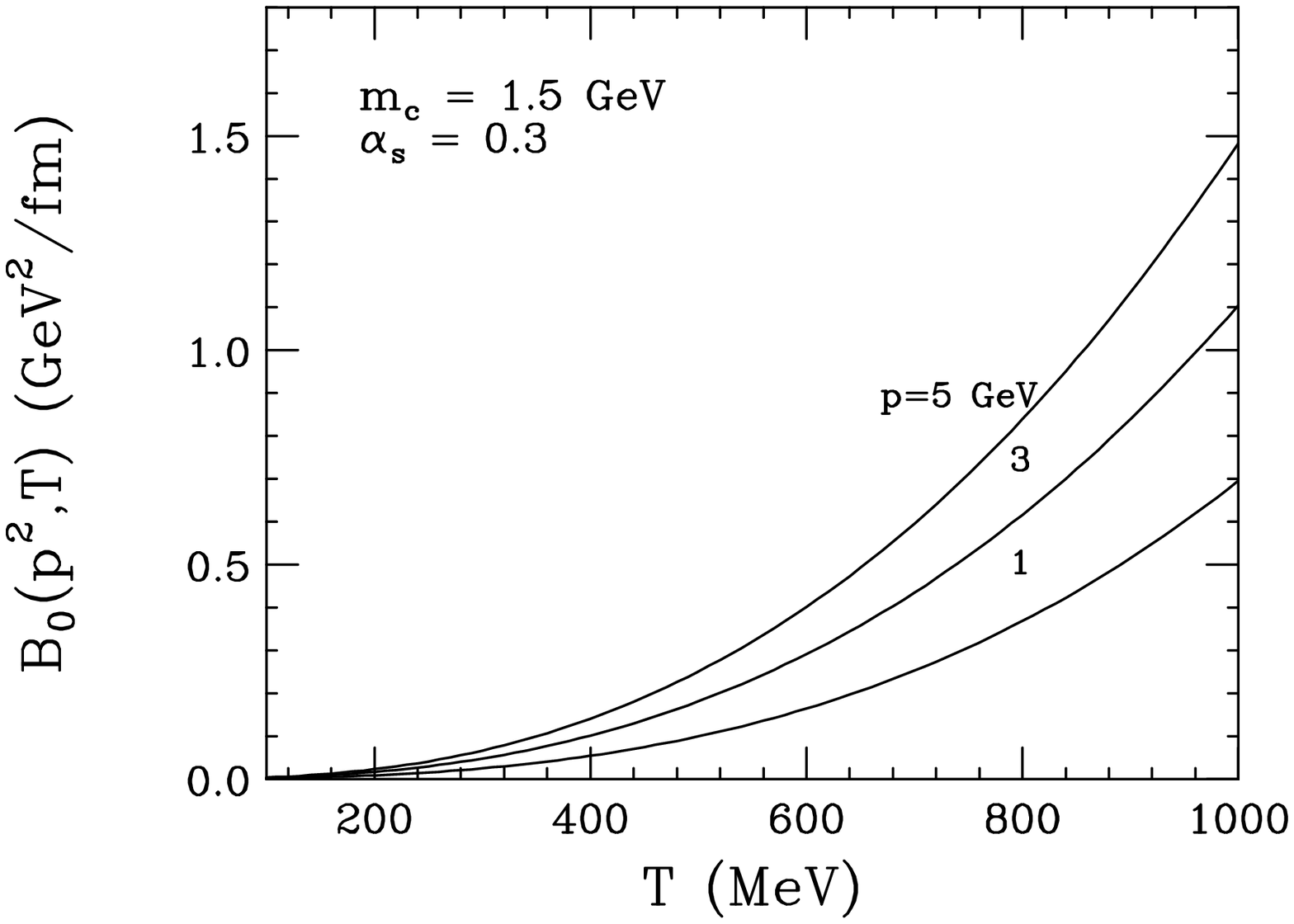}
\vskip 0.1in
\epsfxsize=3.25in
\epsfbox{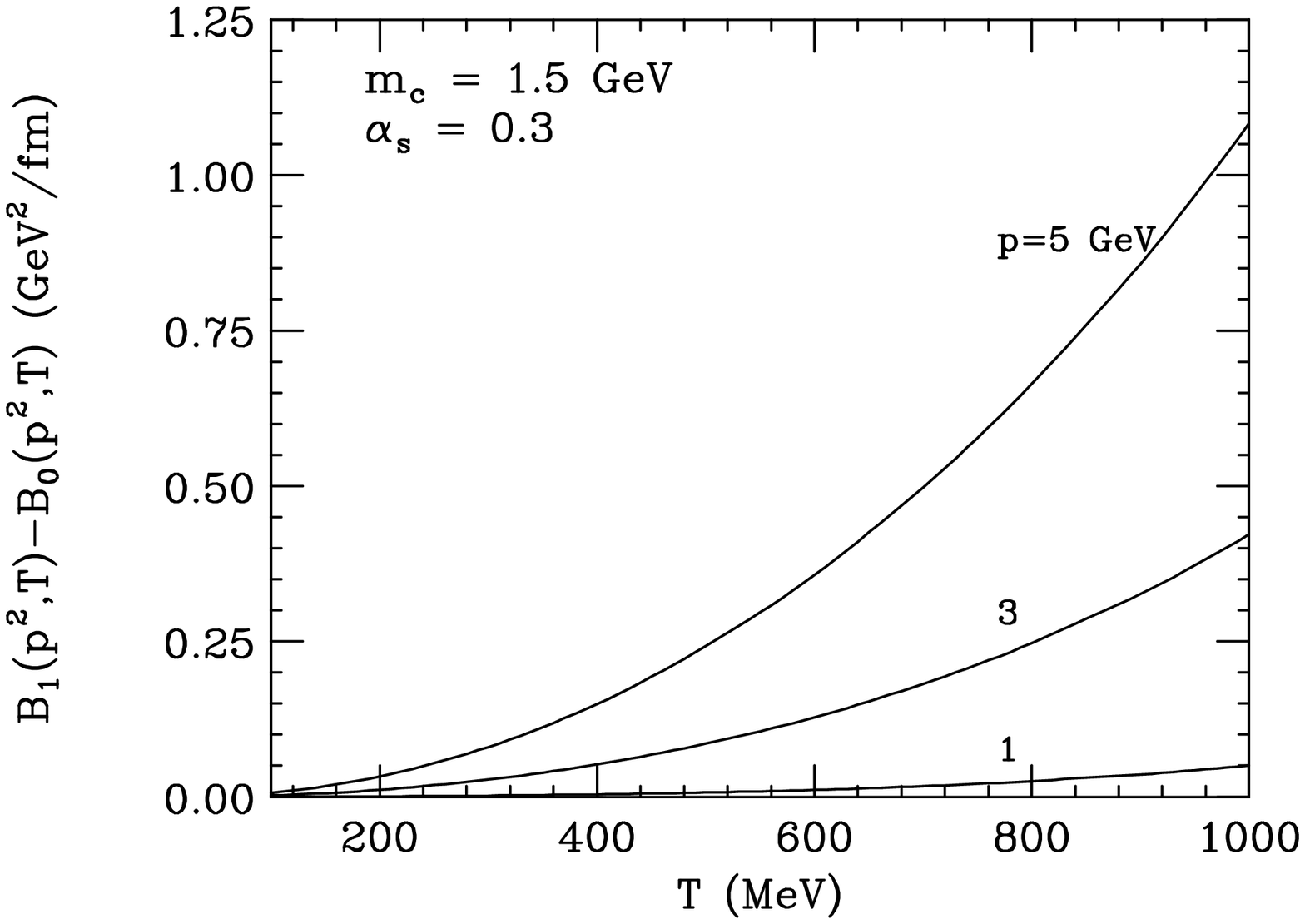}
\vskip 0.1in
\caption{The temperature dependence of the drag ($A$) and the diffusion 
($B_0$ and $B_1-B_0$) coefficients 
for charm quarks in a fully equilibrated quark gluon plasma.
The results shown employ the proper quantum statistics for the gluon and the
quark distributions.}
\end{figure}
The temperature dependence of the drag and
the diffusion coefficients for a fully equilibrated
QGP is shown in Figs.~3(a--c). We see that the drag coefficient drops
rapidly with decline in temperature, so that a low temperature  QGP hardly
offers any resistance to the motion of massive quarks, whose  final momentum
will thus be decided by the time it spends in the very hot plasma.

\subsection{Results for Chemically Equilibrating Plasma at 
RHIC  \& LHC Energies}

The results given so far are for a fully equilibrated plasma. We have already 
discussed that the plasma likely to be created in relativistic heavy ion
collisions is far from chemical equilibrium. In Fig.~4(a) we give the evolution
of the temperature and fugacities for the plasma likely to be created at
RHIC, using the procedure disussed earlier. The appropriate drag
and diffusion coefficients at any time $\tau$ operating on the heavy
quark are obtained by accounting for the current fugacities and the temperature.
In order to avoid confusion, we define 
\begin{equation}
\alpha(p^2,\tau)=A\left( p^2,T (\tau),\lambda_q (\tau), \lambda_g (\tau)\right),
\end{equation}
and similarly $\beta_0=B_0$, $\beta_1=B_1$. The time dependence
of the drag and the diffusion coefficients for some typical values of the
momenta are given in Figs.~4(b--d). We see that the drag
and the diffusion coefficients are large
only at the very early times (due to the high temperatures then) and drop
rapidly as the plasma cools.

The corresponding results for the plasma created at the LHC energies are
given in Figs.~5(a--d). We now see that the drag and the diffusion coefficients
are a factor of $\approx$ 3 larger at any given time for the same momentum
compared to the case at RHIC.
Thus we see that the rapid cooling of the plasma and the small values of the
fugacities ensure that the charm quarks will experience strong drag
force only very early in their life both at the RHIC and the LHC. 
The force is also unlikely to cause a complete stopping and diffusion of the
charm quarks. 
\begin{figure}
\epsfxsize=3.25in
\epsfbox{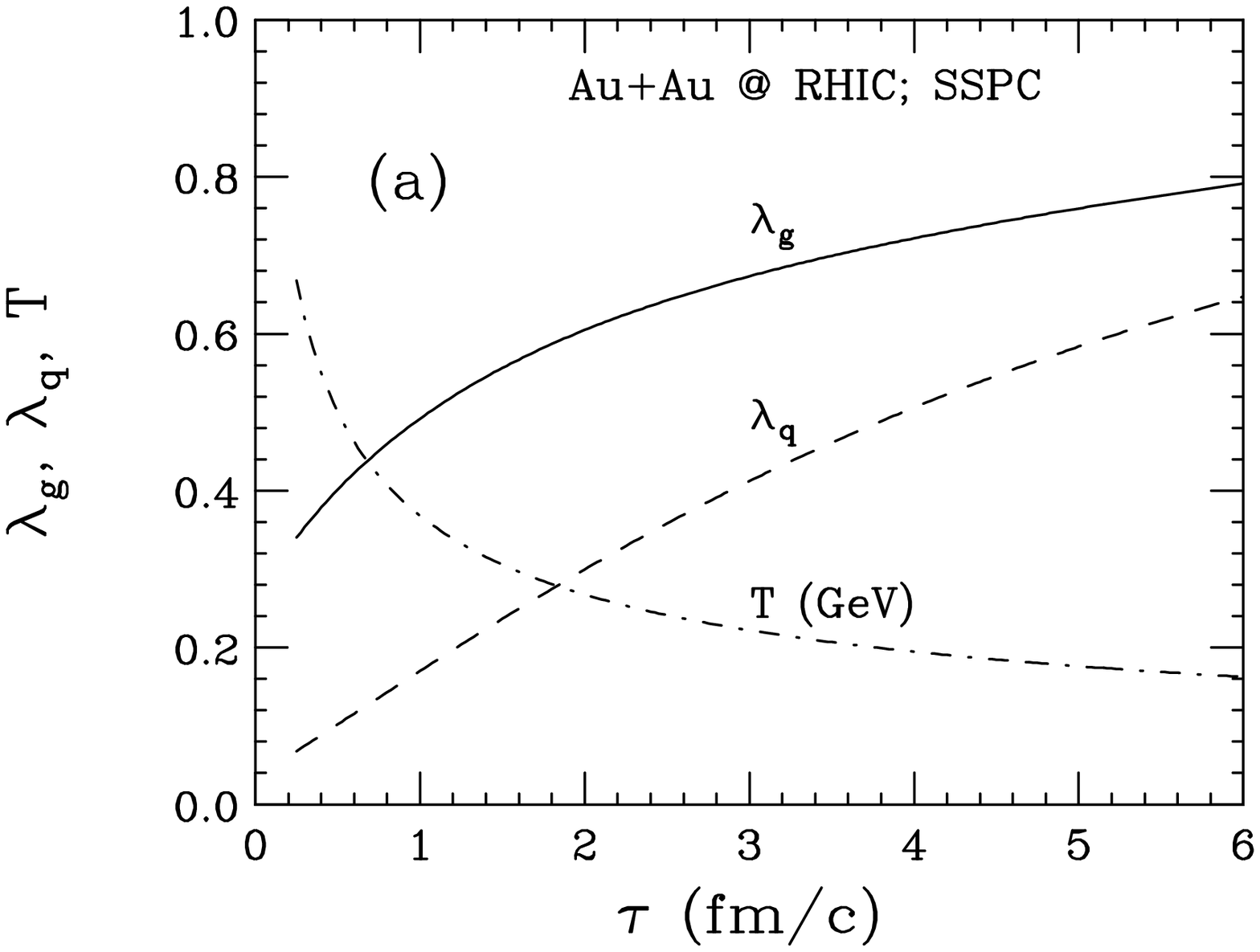}
\vskip 0.2in
\epsfxsize=3.25in
\epsfbox{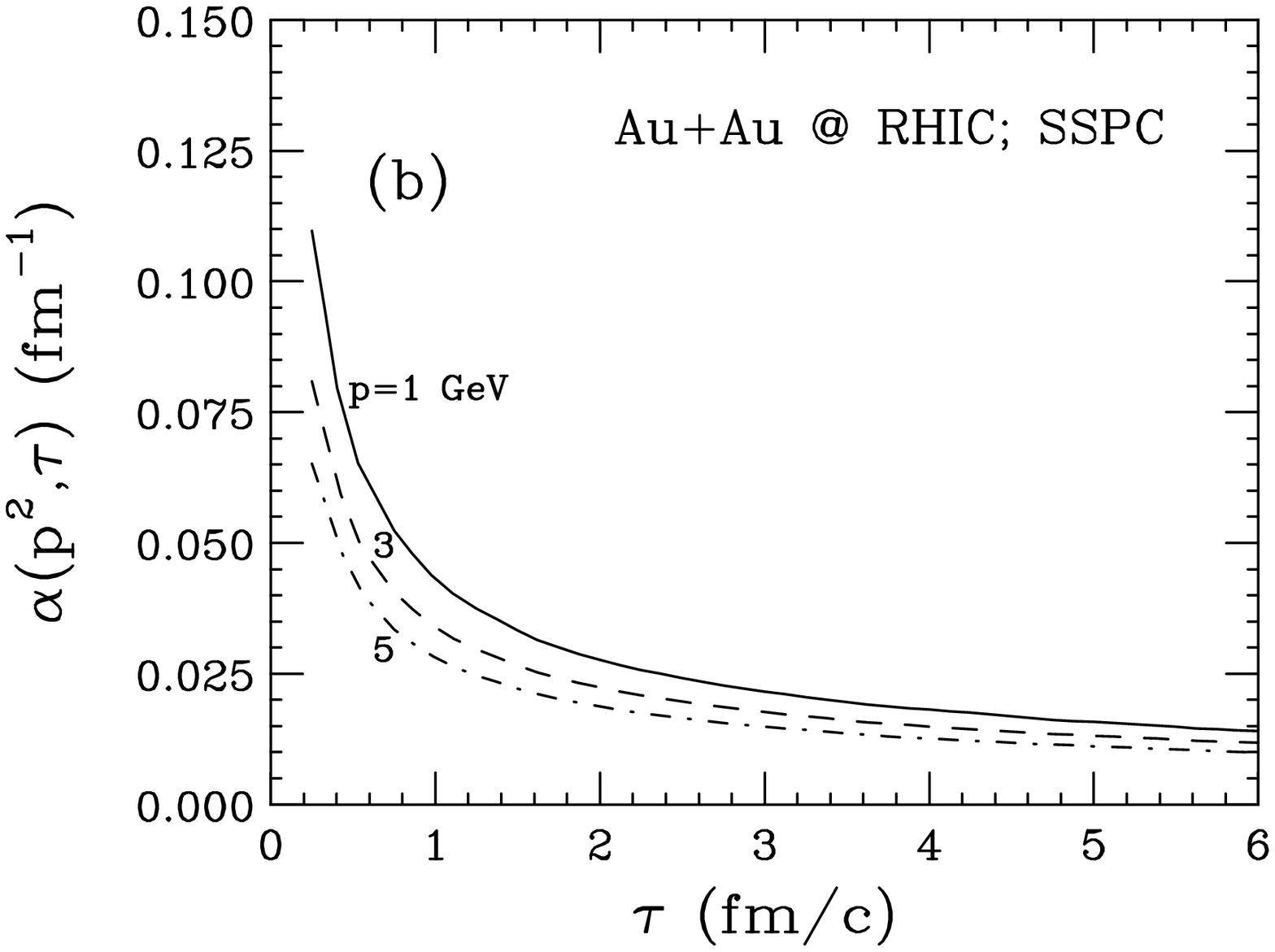}
\vskip 0.2in
\epsfxsize=3.25in
\epsfbox{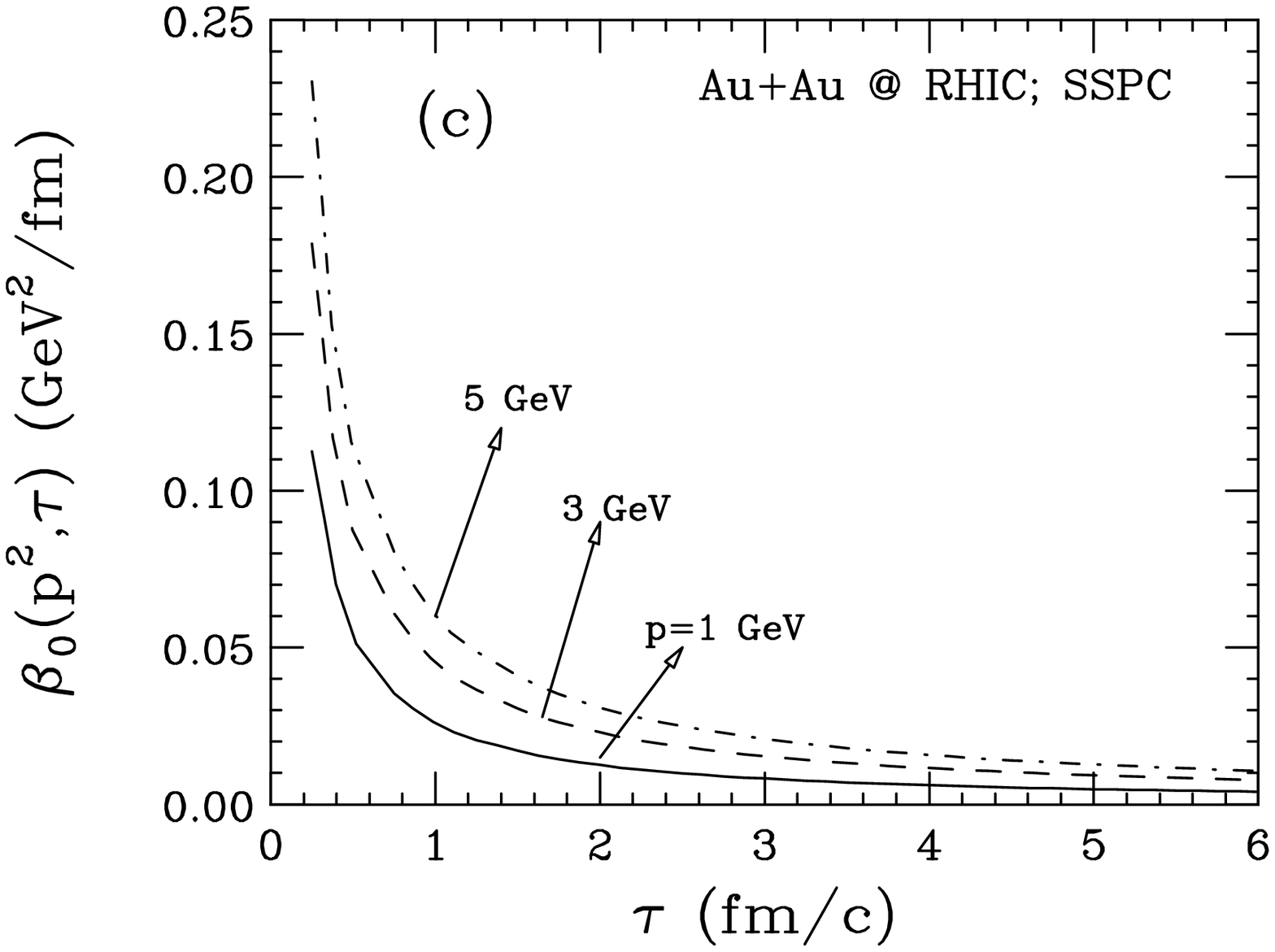}
\vskip 0.2in
\epsfxsize=3.25in
\epsfbox{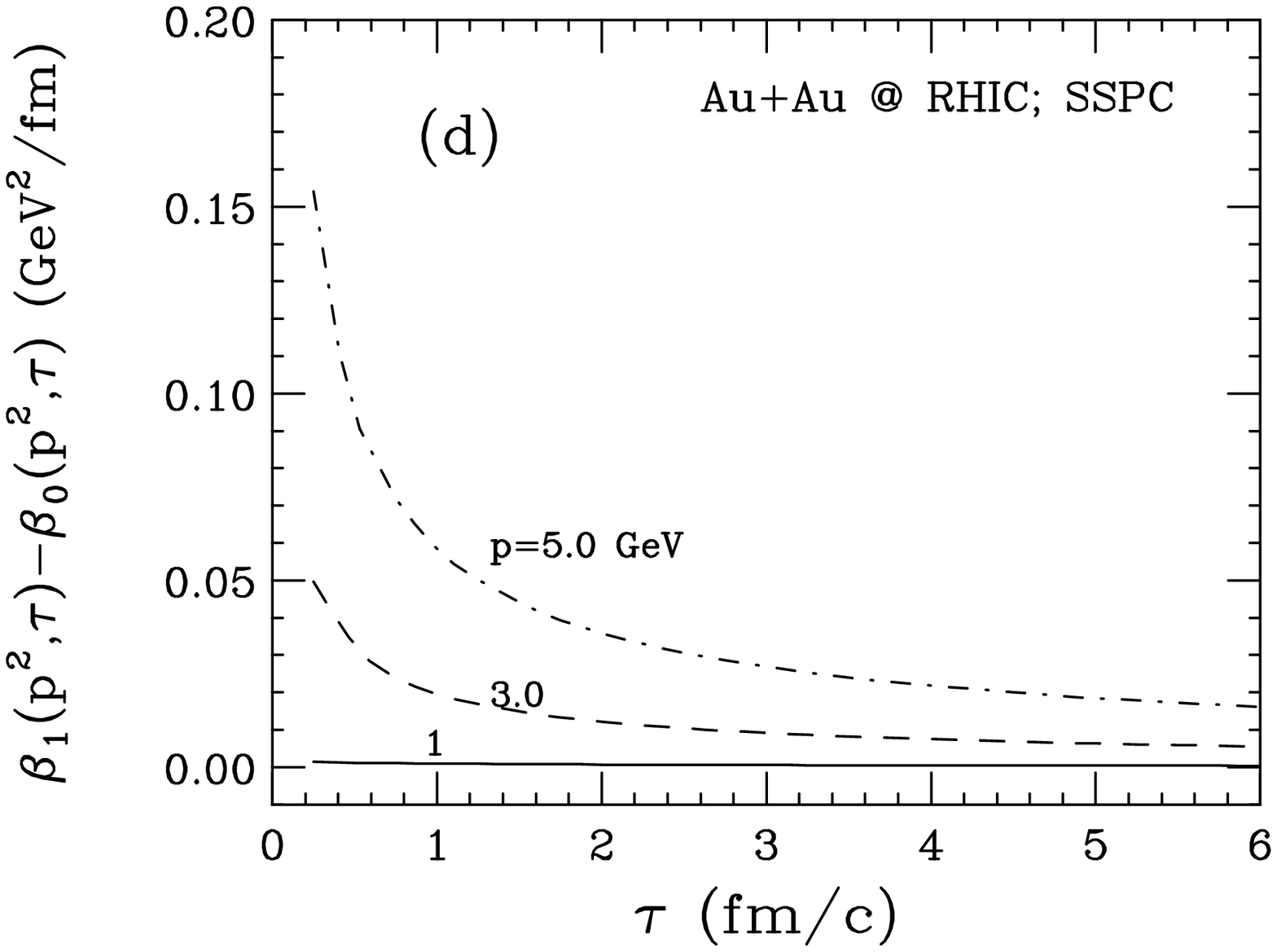}
\vskip 0.3in
\caption{(a)The gluon fugacities ($\lambda_g$), the light quark fugacities 
($\lambda_q$) and the temperature ($T$ in GeV) as a function of proper time 
($\tau$) for a longitudinally expanding plasma likely to be created in 
Au+Au collisions at RHIC. 
(b--d) The variation of the drag ($\alpha$) and the diffusion 
($\beta_0$ and $\beta_1 -\beta_0$) coefficients with proper time ($\tau$) 
for different charm quark momenta.}
\end{figure}

\vskip 0.1 in
The final momentum of the charm quark is likely to
carry the effects of energy loss which in turn is affected by the
cooling as well as the chemical evolution of the plasma. 
Thus for example, {\em if} we had a fully equilibrated QGP at the
same initial temperatures as used here, then the drag at RHIC would
be larger by more than a factor of three while at LHC energies, it
would be large by a factor of more than two. Such large values for the
drag force would ensure that at least the slower charm quarks produced
at the LHC energies would come to rest soon after the creation and  
diffuse later in a manner discussed by Svetitsky.
The final momenta of such charm quark (or D meson) would then
be decided by the hadronization temperature.

\begin{figure}
\epsfxsize=3.25in
\epsfbox{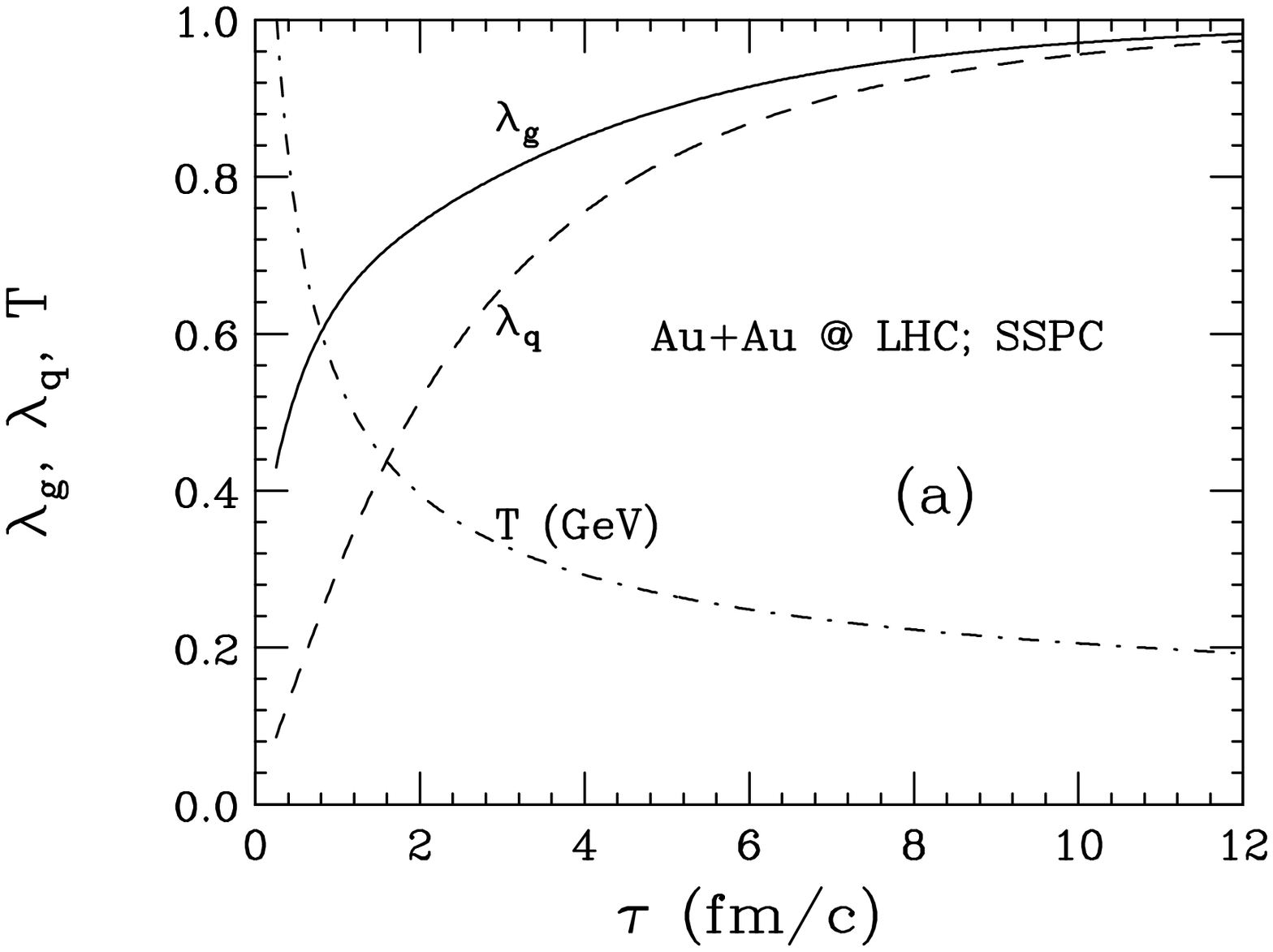}
\vskip 0.2in
\epsfxsize=3.25in
\epsfbox{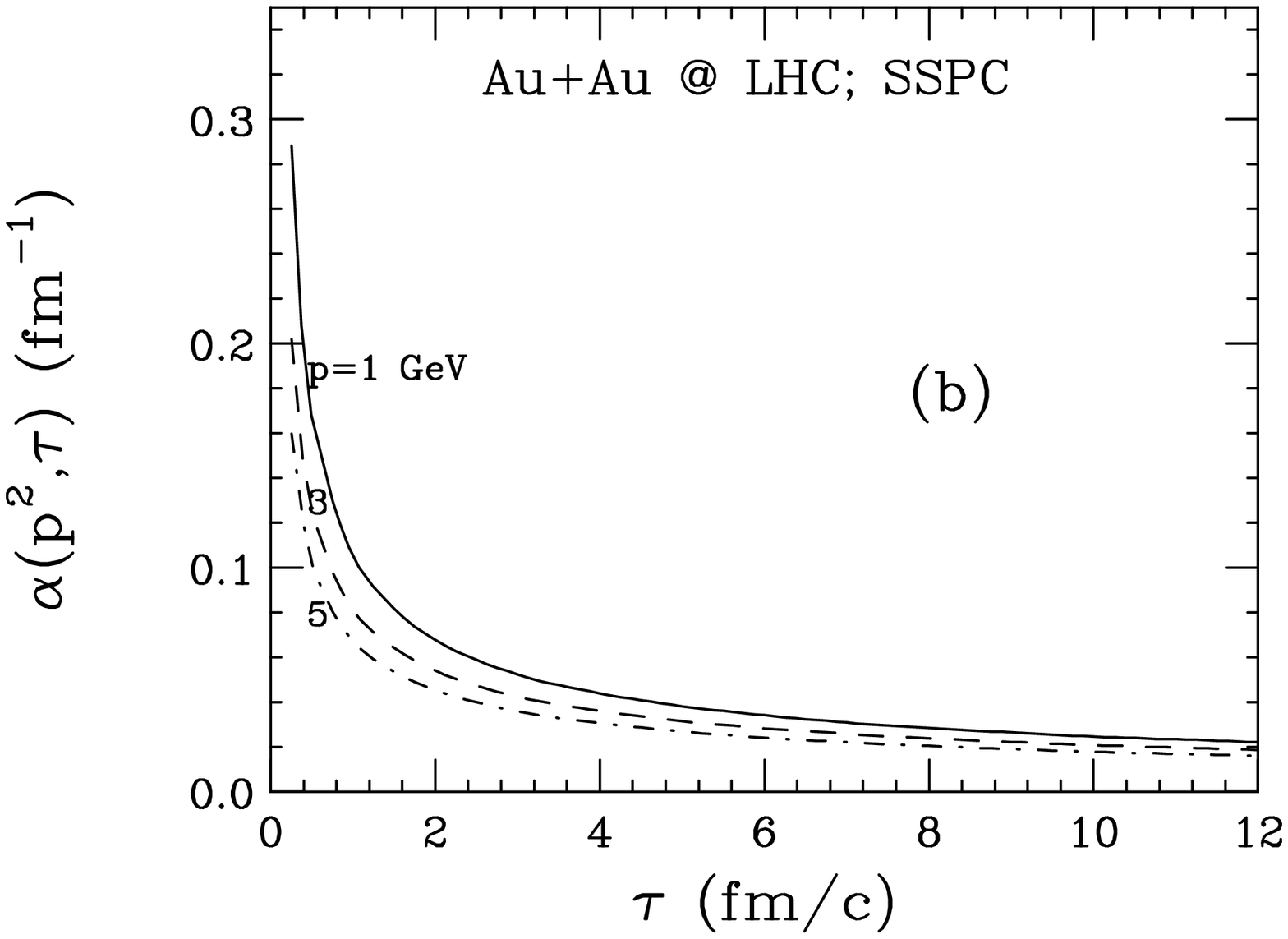}
\vskip 0.2in
\epsfxsize=3.25in
\epsfbox{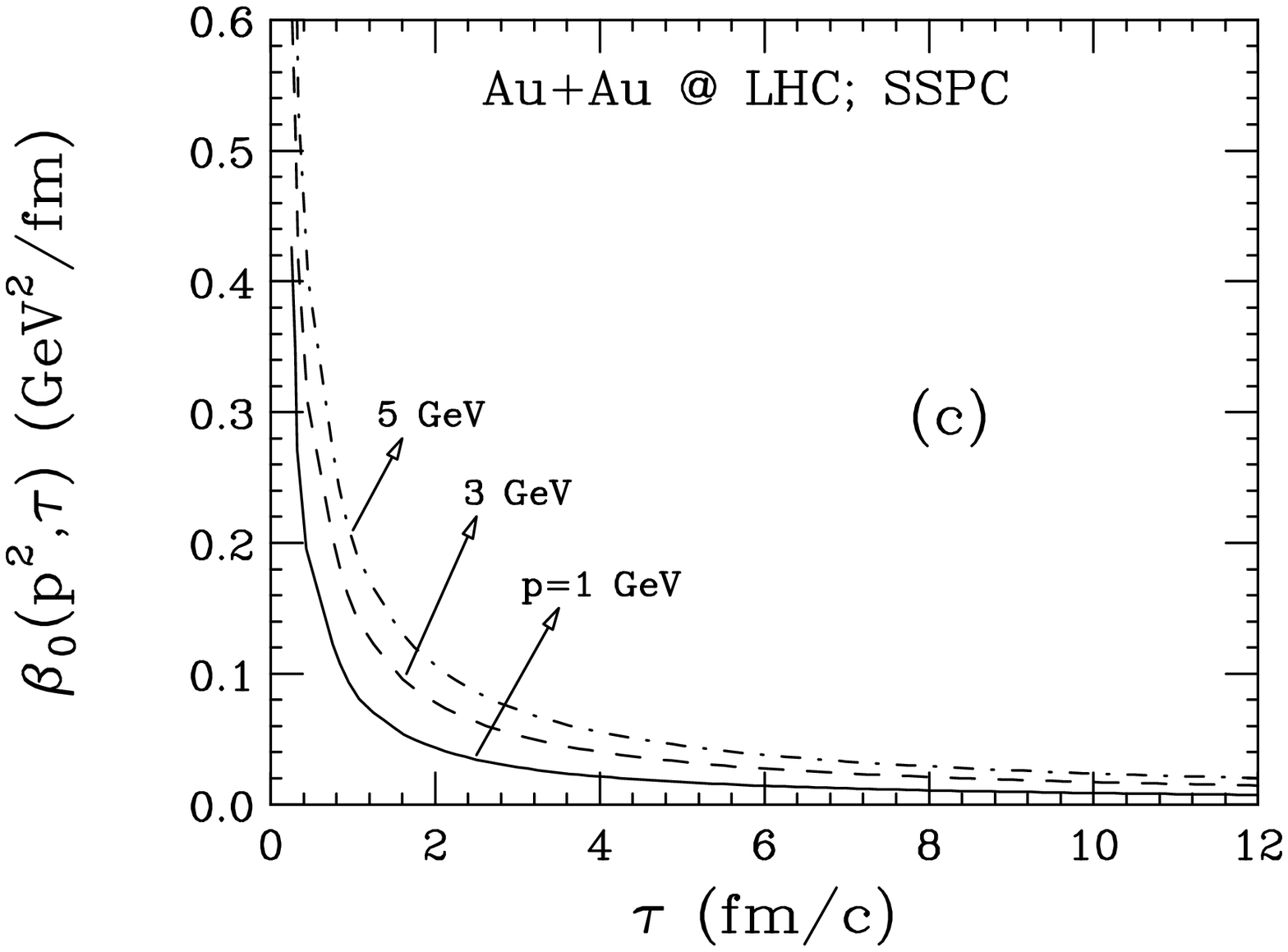}
\vskip 0.2in
\epsfxsize=3.25in
\epsfbox{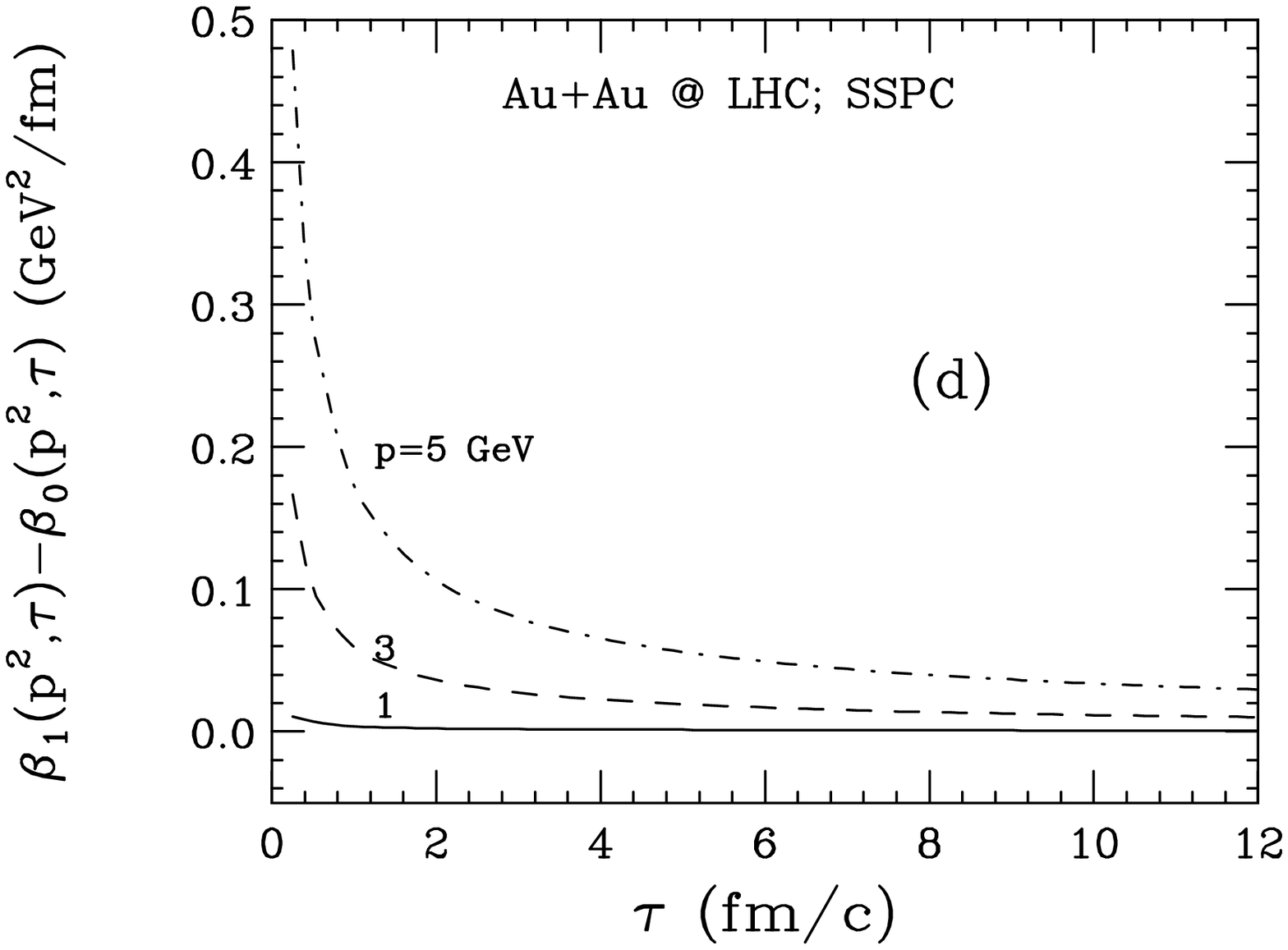}
\vskip 0.3in
\caption{Same as Fig.~4 for LHC energies.}
\end{figure}

In order to get an idea of the energy loss suffered by the charm quarks
produced at RHIC and LHC, we have evaluated their final momentum after
they travel through the plasma for the duration of the QGP phase.

Thus we write
\begin{equation}
\frac{d{\mathbf p}}{d\tau}=-\alpha (p^2,\tau) {\mathbf p}
\end{equation}
and solve for ${\mathbf p}$ as a function of $\tau$. The results shown
in Fig.~6 are very revealing.  We find that by the time the QGP phase is
over, the charm quark which was produced at the beginning of the collision
would have lost up to 40\% of its initial momentum at LHC energies, while
at the RHIC, the energy loss may not exceed 10--12\%.

\begin{figure}
\epsfxsize=3.25in
\epsfbox{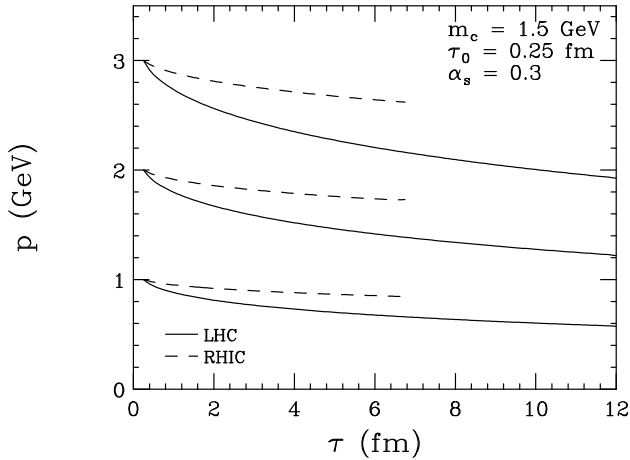}
\vskip 0.3in
\caption{The evolution of the charm quark momentum as a function of 
time in a partially equilibrated plasma at RHIC and LHC. The charm quarks 
are assumed to be produced with momenta of 1, 2, and 3 GeV at the time 
$\tau_0$.}
\end{figure}

\subsection{Distribution of Charm Quarks at RHIC \& LHC}
This has very interesting consequences. At RHIC energies, one may thus
treat charm quark, almost as penetrating probes, which will provide
information about the momentum distribution of the initially produced
charm quark pairs. One can estimate the momentum distribution of
initially produced charm pairs in  a central Au+Au collisions by scaling
the heavy-quark pair production cross-section in pp collisions 
as~\cite{mcgaughey} 
\begin{equation}
{\frac{dN}{d^2p_Tdy}} = {\frac{1}{\pi}} T_{AB}(b=0) 
{\frac{d\sigma}{dp^2_Tdy}} \label{initial}
\end{equation}
where $T_{AB}$ is the nuclear thickness factor. For central Au+Au 
collisions it's value is 293.2 fm$^{-2}$~\cite{evw}. 
One may further approximate the thermal production of the charm quark 
pairs from Eq.(25) of Ref.~\cite{levai}.
The result of this study is shown in Fig.~7. One may obtain the 
``energy-loss corrected'' distribution by decreasing the momenta by about 10\%.
 This also implies that the dileptons from the annihilation
of quarks would continue to remain buried under the leptons originating 
from the open charm decay, as originally inferred by authors of 
Ref.~\cite{ramona}.
We may also add that as the charm quarks do not stop/diffuse at RHIC, they
will not be affected by the transverse velocities which may develop 
during the QGP phase.

The situation is more complex at the LHC energies. Now the charm quarks 
can not be
treated as penetrating probes. The results given
in Fig.~8 correspond to situation when we treat them as penetrating
probes, as in the early studies in this field. As most of
the heavy quarks are produced very early in the collision, we can still
get an idea of the final momentum distribution by decreasing
the momenta in this figure by about 40\%. This however may
not be quite appropriate, as the charm quarks which have lost
a large fraction of their initial momentum due to the drag force
will start getting affected by the transverse flow of the QGP
 which can 
grow to large values at the LHC energies~\cite{munshi}. It is not clear that
this situation can be  accurately handled
 within a hydrodynamic description of the
expansion. A more appropriate description could be the parton cascade model
which includes collisions and even radiations from the charm quark. This
work is under progress.

\begin{figure}
\epsfxsize=3.25in
\epsfbox{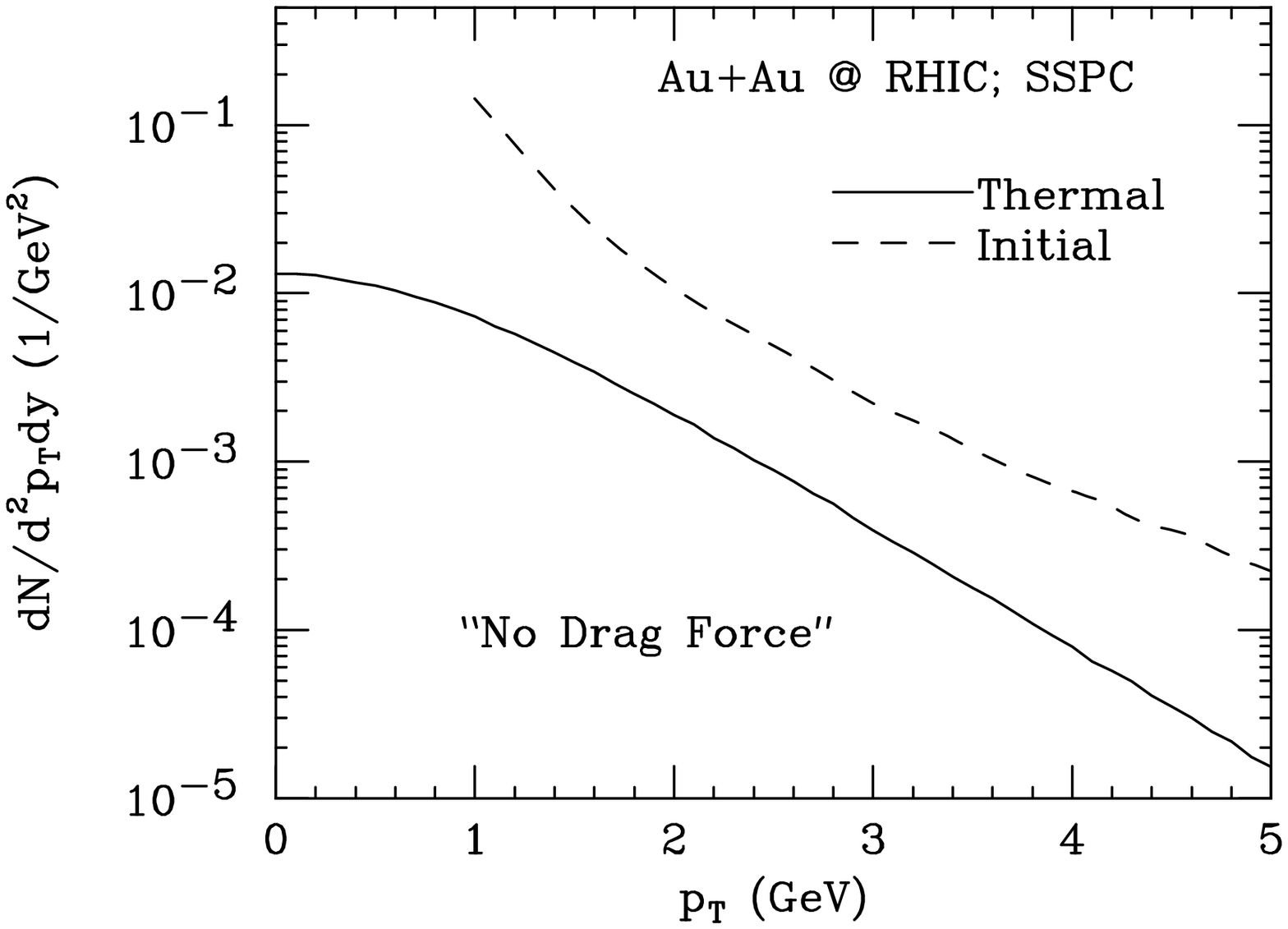}
\vskip 0.3in
\caption{The momentum distribution of charm quarks for RHIC energies {\it 
without energy loss}. }
\vskip 0.3in
\epsfxsize=3.25in
\epsfbox{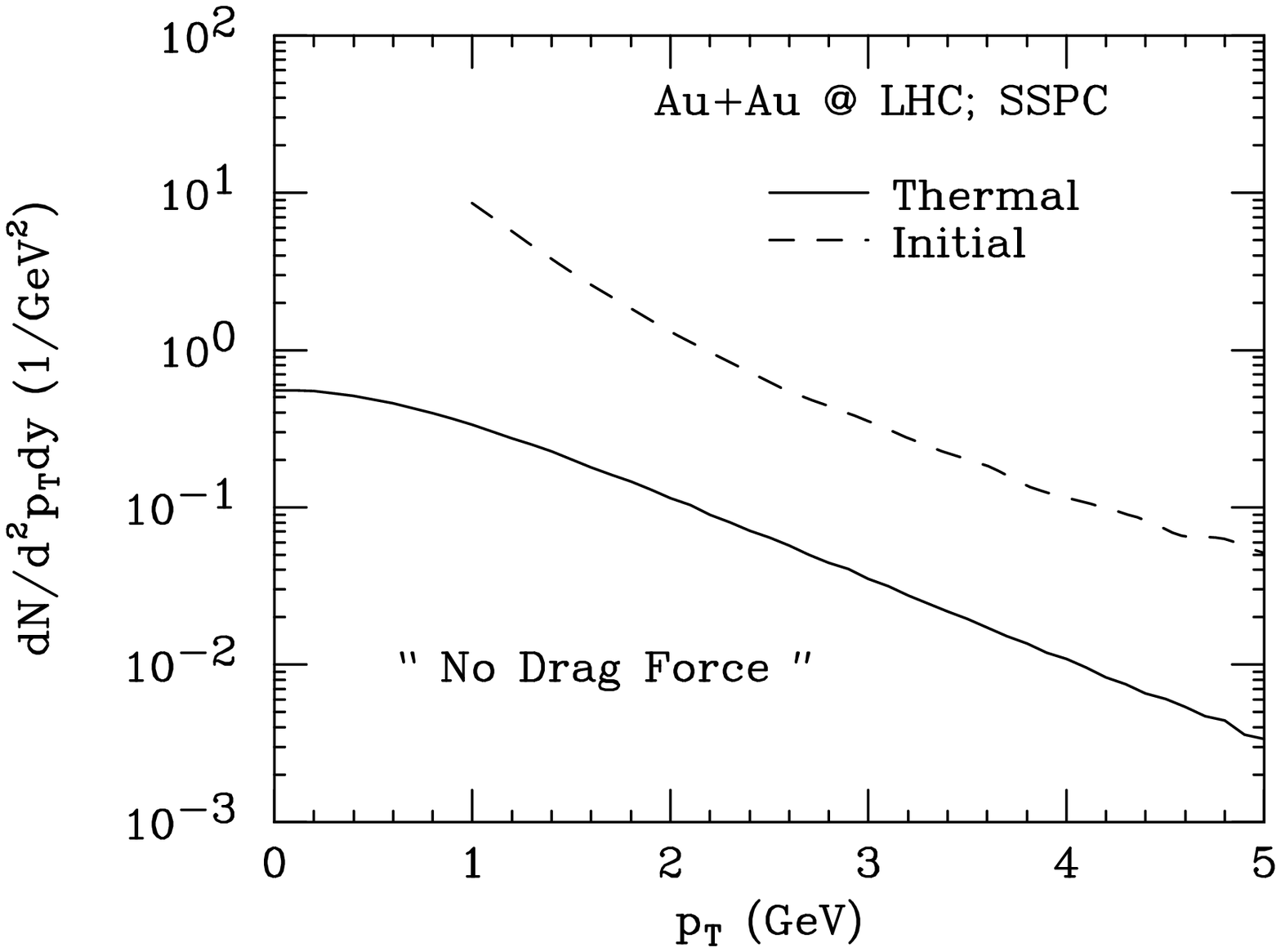}
\vskip 0.3in
\caption{The momentum distribution of charm quarks for LHC energies {\it 
without energy loss}. }
\end{figure}
All the same it is not difficult to imagine that the decrease in the 
slope of the $p_T$ distribution of the charm quarks is
 offset to a large extent  by the flow velocity likely to develop at the LHC
energies, and thus the dileptons from the plasma will perhaps remain
buried under the open charm decay, at these energies as well, unless 
 we look at too large masses.

\section{SUMMARY}

In brief, we have obtained the drag and diffusion coefficients for
charm quarks in a chemically equilibrating quark gluon plasma
 which may be produced at RHIC and LHC energies. Using a set of reasonable
initial conditions, we find that a charm quark produced at the early
stage of the collision may loose up to 10\% of its momentum during the
life-time of the QGP phase at the RHIC energies. This suggests that
at the RHIC energies, dileptons originating from annihilation of quarks
may remain buried under the background from open charm decay. The situation
at LHC is more complex, as the charm quarks may loose up to 40\% of the
initial momentum. This could however be somewhat offset by the transverse
flow of the QGP fluid, which could be large there. Further work,
preferably within a parton cascade model is needed to settle some
of these questions in a more definite manner, though we feel that
even at the LHC energies, the dilepton signal may remain buried under the
background from open charm decay. We may add that some recent
works in this direction have used too large values for the energy loss,
and arrived at results differing from our observations.

The small energy loss seen for heavy quarks in an equilibrating 
plasma could lead to a difference in the ``quenching" of a heavy quark 
jet as compared to that of a gluonic jet. (Radiative loss of gluonic jets 
will presumably not be reduced, considerably, in an unequilibrated plasma.) 
This can be of great interest~\cite{markus1}.
\bigskip

\section*{ACKNOWLEDGEMENTS} We are most grateful to Benjamin Svetitsky for
a very useful correspondence. We acknowledge helpful discussions with
Rudolf Baier, Berndt M\"{u}ller, Bikash Sinha, and Markus Thoma.

\end{document}